\documentclass[a4paper,12pt]{article}

\usepackage{graphics,psfrag}
\usepackage{amsmath,amsthm,amssymb,epsfig,euscript,array,cite,cancel,color}
\usepackage{cases,empheq}
\usepackage[colorlinks=true]{hyperref}
\usepackage{verbatim}
\usepackage{ulem}
   \usepackage[vcentermath]{youngtab}   
   
   \usepackage[usenames,dvipsnames]{xcolor}  

\usepackage[paper=a4paper,dvips,top=2.54cm,left=2.74cm,right=2.34cm,
    foot=1cm,bottom=2.54cm]{geometry}

\theoremstyle{definition} 

\theoremstyle{definition}

\numberwithin{trial}{subsection}

\numberwithin{dtl}{subsection}

\theoremstyle{remark}

\newcounter{multieqs}

\newcommand{\be}{\begin{equation}}
\newcommand{\ee}{\end{equation}}
\newcommand{\eq}[1]{(\ref{#1})}
\newcommand{\bit}{\begin{itemize}}  \newcommand{\eit}{\end{itemize}}

\newcommand{\bm}[1]{\mbox{\boldmath $#1$}}
\newcommand{\rf}[1]{(\ref{#1})}

\def\bd{\begin{document}}
\def\ed{\end{document}}
\def\nn{\nonumber}
\def\bea{\begin{eqnarray}}
\def\eea{\end{eqnarray}}
\let\bm=\bibitem

\def\la{\langle}
\def\ra{\rangle}

\def\npb#1#2#3{Nucl. Phys. {\bf{B#1}} #3 (#2)}
\def\plb#1#2#3{Phys. Lett. {\bf{#1B}} #3 (#2)}
\def\prl#1#2#3{Phys. Rev. Lett. {\bf{#1}} #3 (#2)}
\def\prd#1#2#3{Phys. Rev. {D \bf{#1}} #3 (#2)}
\def\cmp#1#2#3{Comm. Math. Phys. {\bf{#1}} #3 (#2)}
\def\cqg#1#2#3{Class. Quantum Grav. {\bf{#1}} #3 (#2)}
\def\nppsa#1#2#3{Nucl. Phys. B (Proc. Suppl.) {\bf{#1A}}#3 (#2)}
\def\ap#1#2#3{Ann. of Phys. {\bf{#1}} #3 (#2)}
\def\ijmp#1#2#3{Int. J. Mod. Phys. {\bf{A#1}} #3 (#2)}
\def\rmp#1#2#3{Rev. Mod. Phys. {\bf{#1}} #3 (#2)}
\def\mpla#1#2#3{Mod. Phys. Lett. {\bf A#1} #3 (#2)}
\def\jhep#1#2#3{J. High Energy Phys. {\bf #1} #3 (#2)}
\def\atmp#1#2#3{Adv. Theor. Math. Phys. {\bf #1} #3 (#2)}

\def\N{{\cal N}}
\def\sst{\scriptscriptstyle}
\def\thetabar{\bar\theta}
\def\Tr{{\rm Tr}}
\def\one{\mbox{1 \kern-.59em {\rm l}}}

\def\a{\alpha}      \def\da{{\dot\alpha}}  \def\dA{{\dot A}}
\def\b{\beta}       \def\db{{\dot\beta}}  
\def\g{\gamma}  \def\G{\Gamma}  \def\dc{{\dot\gamma}}  
\def\d{\delta}  \def\D{\Delta}  \def\ddt{\dot\delta}  
\def\e{\epsilon}        \def\ve{\varepsilon}  
\def\f{\phi}    \def\F{\Phi}    \def\vvf{\f}  
\def\h{\eta}  
\def\k{\kappa}  
\def\l{{\lambda}} \def\L{\Lambda}  
\def\m{\mu} \def\n{\nu}  
\def\om{\omega}  
\def\p{\pi} \def\P{\Pi}  
\def\r{\rho}  
\def\s{\sigma}  \def\S{\Sigma}  
\def\t{\tau}  
\def\th{\theta} \def\Th{\Theta} \def\vth{\vartheta}  
\def\X{\Xeta}  
\def\z{\zeta}  

\def\na{\nabla}  

\def\cA{{\cal A}} \def\cB{{\cal B}} \def\cC{{\cal C}}  
\def\cD{{\cal D}} \def\cE{{\cal E}} \def\cF{{\cal F}}  
\def\cG{{\cal G}} \def\cH{{\cal H}} \def\cI{{\cal I}}  
\def\cJ{{\cal J}} \def\cK{{\cal K}} \def\cL{{\cal L}}  
\def\cM{{\cal M}} \def\cN{{\cal N}} \def\cO{{\cal O}}  
\def\cP{{\cal P}} \def\cQ{{\cal Q}} \def\cR{{\cal R}}  
\def\cS{{\cal S}} \def\cT{{\cal T}} \def\cU{{\cal U}}  
\def\cV{{\cal V}} \def\cW{{\cal W}} \def\cX{{\cal X}}  
\def\cY{{\cal Y}} \def\cZ{{\cal Z}}

\def\ua{{\underline{\alpha}}} 
 \def\ub{\underline{\phantom{\alpha}}\!\!\!\beta}  
\def\uc{\underline{\phantom{\alpha}}\!\!\!\gamma}  
\def\um {{\underline{\mu}}}  
\def\ud{{\underline{\delta}}} 
\def\ue{\underline\epsilon}  
\def\una{{\underline a}}\def\uA{{\underline A}}  
\def\unb{{\underline b}}\def\uB{{\underline B}} 
\def\unc{{\underline c}}\def\uC{{\underline C}}  
\def\und{{\underline d}}\def\uD{{\underline D}}  
\def\une{{\underline e}}\def\uE{{\underline E}}  
\def\unf{{\underline{\phantom{e}}\!\!\!\! f}}\def\uF{{\underline F}}  
\def\unm{{\underline m}\def\uM{\underline M}} 
\def\unn{{\underline n}\def\uN{\underline N}} 
\def\unp{{\underline{\phantom{a}}\!\!\! p}}\def\uP{{\underline P}}  
\def\unq{{\underline{\phantom{a}}\!\!\! q}}  
\def\uQ{{\underline{\phantom{A}}\!\!\!\! Q}}  
\def\uH{{\underline{H}}}  
\def\uM{{\underline{M}}}
\def\uN{{\underline{N}}}
  
\def\As {{A \hspace{-6.4pt} \slash}\;}  
\def\bs {{b \hspace{-6.4pt} \slash}\;}  
\def\Ds {{D \hspace{-6.4pt} \slash}\;}
\def\Gts {{\Gt \hspace{-6.4pt} \slash}\;}
\def\ds {{\del \hspace{-6.4pt} \slash}\;}  
\def\ss {{\s \hspace{-6.4pt} \slash}\;}  
\def\ks {{ k \hspace{-6.4pt} \slash}\;}  
\def\ps {{p \hspace{-6.4pt} \slash}\;}   
\def\xs {{x \hspace{-6.4pt} \slash}\;}  
\def\pas {{{p_1} \hspace{-6.4pt} \slash}\;}  
\def\pbs {{{p_2} \hspace{-6.4pt} \slash}\;}   
\def\cFs {{{\cal F} \hspace{-6.4pt} \slash}\;}

\def\Ah{{\hat{A}}}  
\def\Dh{{\hat{D}}}
\def\Gh{{\hat{G}}}
\def\Fh{{\hat{F}}}
\def\Ih{{\hat{I}}} 
\def\Jh{{\hat{J}}} 
\def\Kh{{\hat{K}}}
\def\Lh{{\hat{L}}} 
\def\Ph{{\hat{P}}}
\def\Rh{{\hat{R}}}
\def\Vh{{\hat{V}}} 
\def\Xh{{\hat{X}}}
 
\def\ah{{\hat{a}}}
\def\bh{{\hat{b}}}
\def\ch{{\hat{c}}}
\def\gh{{\hat{g}}}
\def\dh{{\hat{d}}}
\def\hh{{\hat{h}}}
\def\uh{{\hat{u}}}  
\def\vh{{\hat{v}}}
\def\xh{{\hat{x}}} 
\def\yh{{\hat{y}}}
\def\zh{{\hat{z}}}
\def\ph{{\hat{p}}}
\def\qh{{\hat{q}}}
\def\thh{{\hat{t}}}  
\def\xih{\hat{\xi}}  
\def\Psih{\hat{\Psi}}    
\def\mh{{\hat{m}}}
\def\nh{{\hat{n}}}
\def\ih{{\hat{i}}}
\def\jh{{\hat{j}}}
\def\kh{{\hat{k}}}
\def\aah{{\hat{\alpha}}}
\def\bbh{{\hat{\beta}}}
\def\ggh{{\hat{\gamma}}}
\def\llh{{\hat{\ell}}} 
\def\ph{{\hat{p}}}

\def\psit{\tilde{\psi}}  
\def\Psit{\tilde{\Psi}}   
\def\Psibt{\tilde{\bar{Psi}}}  

\def\st{\tilde{\sigma}}  

\def\delt{\tilde{\delta}}
\def\Phit{\tilde{\Phi}}   
\def\Phitb{\overline{\tilde{Phi}}}  
\def\tht{\tilde{\th}}  
\def\lt{\tilde{\l}}
\def\chit{\tilde{\chi}}   
\def\phit{\tilde{\phi}} 

\def\At{\tilde{A}}
\def\Bt{\tilde{B}}
\def\Ct{\tilde{C}}
\def\Dt{\tilde{D}}
\def\Et{\tilde{E}}
\def\Ft{\tilde{F}}
\def\Gt{\tilde{G}}
\def\Ht{\tilde{H}}
\def\It{\tilde{I}}
\def\Jt{\tilde{J}}
\def\Qt{\tilde{Q}}  
\def\Rt{\tilde{R}}  
\def\Mt{\tilde{M }}  
\def\Nt{\tilde{N}}   
\def\St{\tilde{S}}
\def\Vt{\tilde{V}}
\def\Xt{\tilde{X}} 
\def\at{\tilde{a}}
\def\ct{\tilde{c}}
\def\dt{\tilde{d}}
\def\htt{\tilde{h}} 
\def\ft{\tilde{f}}
\def\gt{\tilde{g}}
\def\pt{\tilde{p}}  
\def\qt{\tilde{q}}  
\def\vt{\tilde{v}}  
\def\nt{\tilde{n}}  
\def\ut{\tilde{u}}  
\def\wt{\tilde{w}}  
\def\zt{\tilde{z}} 
\def\xt{\tilde{x}} 
\def\yt{\tilde{y}} 
\def\Psit{\tilde{\Psi}}
\def\vphit{\tilde{\varphi}}

\def\gamt{\tilde{\gamma}}

\def\Tt{\tilde{T}}

\def\ot{{\tilde{\omega}}}

\def\eb{\bar{\epsilon}} 
\def\delb{\bar{\partial}}  
\def\thb{\bar{\theta}}
\def\Thb{{\bar{\Theta}}}
\def\mub{\bar{\mu}}
\def\lamb{\bar{\l}}
\def\psib{\bar{\psi}}
\def\sb{\bar{\sigma}}
\def\xib{\bar{\xi}}
\def\chib{\bar{\chi}}

\def\Psib{\bar{\Psi}}
\def\Phib{\bar{\Phi}}
\def\Lamb{\bar{\Lambda}}
\def\Sb{{\overline \Sigma}}
\def\cb{\bar{c}}
\def\hb{\bar{h}}
\def\qb{\bar{q}}
\def\wb{\bar{w}}
\def\zb{{\bar{z}}}
\def\Hb{\bar{H}}
\def\Qb{{\bar Q}}
\def\Omegab{\overline{\Omega}}
\def\ob{\overline{\omega}}
\def\Gab{{\bar{\Gamma}}}

\def\Ab{{\overline A}} \def\Bb{{\overline B}} \def\Cb{{\overline C}}  
\def\Db{{\overline D}} \def\Eb{{\overline E}} \def\Fb{{\overline F}}  
\def\Gb{{\overline G}} 
\def\Ib{{\overline I}}  
\def\Jb{{\overline J}} \def\Kb{{\overline K}} \def\Lb{{\overline L}}  
\def\Mb{{\overline M}} \def\Nb{{\overline N}} \def\Ob{{\overline O}}  
\def\Pb{{\overline P}}  \def\Rb{{\overline R}}  
 \def\Tb{{\overline T}} \def\Ub{{\overline U}}  
\def\Vb{{\overline V}} \def\Wb{{\overline W}} \def\Xb{{\overline X}}  
\def\Yb{{\overline Y}} \def\Zb{{\overline Z}}  

\def\fb{{\overline f}}
\def\gb{{\overline g}}
\def\mb{{\overline m}}
\def\lb{{\overline l}}
\def\yb{{\overline y}}
  
\def\ldel{{\overleftarrow{\del}}}
\def\rdel{{\overrightarrow{\del}}}
\def\ldeldel{{\overleftarrow{\del^2}}}
\def\rdeldel{{\overrightarrow{\del^2}}}
\def\ldelb{{\overleftarrow{\bar{\del}}}}
\def\rdelb{{\overrightarrow{\bar{\del}}}}

\def\ba{{\bf a}} 
\def\bk{{\bf k}}  
\def\bl{{\bf l}}  
\def\bp{{\bf p}}  
\def\bq{{\bf q}}  
\def\br{{\bf r}}
\def\bt{{\bf t}}
\def\bu{{\bf u}}
\def\bv{{\bf v}}
\def\bx{{\bf x}}  
\def\by{{\bf y}}  
\def\bR{{\bf R}}  
\def\bV{{\bf V}}  
\def\bK{{\bf K}}

\def\bone{{\bf 1}}  

\def\va{{\vec a}}
\def\vk{{\vec k}}
\def\vp{{\vec p}}
\def\vq{{\vec q}}
\def\vx{{\vec x}}
\def\vy{{\vec y}}
\def\vu{{\vec u}}
\def\vv{{\vec v}}

\def\vs{{\vec \sigma}}
\def\vtau{{\vec \tau}}

\newcommand{\ov}[1]{\overrightarrow{#1}}

\def\frA{\mathfrak{A}}
\def\frB{\mathfrak{B}}
\def\frC{\mathfrak{C}}
\def\frD{\mathfrak{D}}
\def\frE{\mathfrak{E}}
\def\frF{\mathfrak{F}}
\def\frG{\mathfrak{G}}
\def\frH{\mathfrak{H}}
\def\frM{\mathfrak{M}}
\def\frN{\mathfrak{N}}
\def\frR{\mathfrak{R}}
\def\frW{\mathfrak{W}}

\def\fra{\mathfrak{a}}
\def\frb{\mathfrak{b}}
\def\frf{\mathfrak{f}}
\def\frg{\mathfrak{g}}
\def\frh{\mathfrak{h}}
\def\frl{\mathfrak{l}}
\def\frs{\mathfrak{s}}
\def\fri{\mathfrak{i}}
\def\frj{\mathfrak{j}}

\def\ma{\mathfrak{a}}
\def\mg{\mathfrak{g}}
\def\mR{\mathfrak{R}}
\def\mN{\mathfrak{N}}

\def\d{\delta}\def\D{\Delta}\def\ddt{\dot\delta}  
  
\def\pa{\partial} \def\del{\partial}  
\def\xx{\times}  
\def\uno{\mbox{1 \kern-.59em {\rm l}}}    
  
\def\trp{^{\top}}  
\def\inv{^{-1}}  
\def\dag{{^{\dagger}}}  
\def\pr{^{\prime}}  
  
\def\rar{\rightarrow}  
\def\lar{\leftarrow}  
\def\lrar{\leftrightarrow}  
  
\newcommand{\0}{\,\!}      
\def\one{1\!\!1\,\,}  
\def\im{\imath}  
\def\jm{\jmath}  
  
\newcommand{\tr}{\mbox{tr}}  
\newcommand{\slsh}[1]{/ \!\!\!\! #1}  
  
\def\vac{|0\rangle}  
\def\lvac{\langle 0|}  
  
\def\hlf{\frac{1}{2}}  
\def\ove#1{\frac{1}{#1}}  

\def\Box{\square}  
\def\CC {\mathbb{C}}
\def\FF {\mathbb{F}}
\def\RR{\mathbb{R}}
\def\NN{\mathbb{N}}  
\def\ZZ{\mathbb{Z}}  
\def\bb#1{{\bf #1}}  
\def\bcomment#1{}  
\def\bfhat#1{{\bf \hat{#1}}}  
\def\VEV#1{\left\langle #1\right\rangle}  

\newcommand{\ex}[1]{{\rm e}^{#1}} \def\ii{{\rm i}}  

\newcommand{\lrbrk}[1]{\left(#1\right)}
\newcommand{\lrsbrk}[1]{\left[#1\right]}
\newcommand{\lrcbrk}[1]{\left\{#1\right\}}
\newcommand{\sfrac}[2]{{\textstyle\frac{#1}{#2}}}
 
\def\stw{{\sqrt{2}}}

\def\rf {{\rm f}}
\def\ri {{\rm i}}
\def\rj {{\rm j}}
\def\rk {{\rm k}}
\def\rl {{\rm l}}
\def\rs {{\scriptscriptstyle \rm S}}
\def\rt {{\scriptscriptstyle \rm T}}

\def\rQ {{\scriptscriptstyle \rm \cQ}}
\def\rR {{\scriptscriptstyle \rm \cR}}

\def\cQb{{\cal \Qb}}
\def\cRb{{\cal \Rb}}
\def\cWb{{\cal \Wb}}

\def\fd {{\rm N}}
\def\afd {{\overline{\rm N}}}

\def \II {I\hspace{-.1em}I\hspace{.1em}}
\def \IIA {\mbox{\II A\hspace{.2em}}}
\def \IIB {\mbox{\II B\hspace{.2em}}}
\def \gs {g^s}
\def \ls {\lambda^s}

\def \I {{\cal I}}
\def \qs {q\hspace{-.53em}/\hspace{.15em}}
\def \ks {k\hspace{-.53em}/\hspace{.15em}}
\def \YM {{\mbox{\tiny YM}}}
\def \gym {g_{\YM}}

\def \Lc {\L_c}
\def\IR{\relax{\rm I\kern-.18em R}}
\def \id {{\bf 1}}

\def\cci{\ell}
\def\ccj{\ell'}

\def \thbb{\overline{\th\th}}
\newcommand \ol{\overline}
\def \lamb{\bar{\lambda}}
\def \vphi{\varphi}
\def \lambh{\hat{\bar{\lambda}}}
\def \lh{\hat{\lambda}}
\def \dd{\ddagger}

\def \ad {\dot{a}}
\def \bd {\dot{b}}
\def \cd {\dot{c}}
\def  \ddd {\dot{d}}
\def \ed {\dot{e}}
\def \fd {\dot{f}}
\def \Bh {\hat{B}}
\def \zm {{(0)}}
\def \nz {{(\text{KK})}}
\def \3{{(3)}}
\def \diag {\text{diag}}
\def \inm {{(m^{-1})}}
\def \3{{(3)}} 
\def \6{{(6)}}
\def \2{{(2)}}
\def \7{{(7)}} 
\def \4{{(4)}}
\def\1{{(1)}}
\def\5{{(5)}}
\def\0{{(0)}}

\def\eh{{\hat{e}}}
\def\fh{{\hat{f}}}
\def\lh{{\hat{l}}}
\def\rh{{\hat{r}}}
\def\wh{{\hat{w}}}
\renewcommand{\mh}{{\hat{m}}}

\def \DBI{{\text{DBI}}}
\def\et{{\tilde{\e}}}
\def\w{{\wedge}}
\def\bbV{{\mathbb{V}}}
\def\M{{(\text{M})}}
\def\T{{(\text{T})}}

\def\Hbt{{\tilde{\bar{H}}}}
\def\Fbt{{\tilde{\bar{F}}}}

\def\fR{{\mathfrak{R}}}
\def\fg{{\mathfrak{g}}}
\def \sk {\textsc{k}}
\def\S{{\Sigma}}
\def\nb{{\nabla}}
\def\bB{{\bf B}}

\colorlet{1}{red}
\colorlet{2}{green}
\colorlet{3}{blue}
\colorlet{4}{cyan}
\def\bJ{{\bold J}}
\def\tR{{\text{(R)}}}
\def\bQ{{\bold Q}}
\def\bo{{\pmb{\omega}}}
\def\ADM{{\text{ADM}}}
\def\st{{\tilde\sigma}}
\def\Fbh{{\Phi_{\text{bh}}}}
\def\bC{{\bold C}}
\def\bep{{\pmb\epsilon}}
\def\EM{{(\text{EM})}}
\def \bTh {\bold \Theta}
\def\bL{{\bold L}}
\def\tG{{\tilde{\Gamma}}}
\def \vL {\mathfrak{L}}
\def\bE{{\bold E}}
\def\nbs{{\overset{*}{\nb}}}
\def\nbt{{\tilde{\nb}}}

\author{Sheng-Lan Ko\footnote{sheng-lank@nu.ac.th}$~^{a,b}$,
 Feng-Li Lin\footnote{fengli.lin@gmail.com}$~^b$ and Bo Ning\footnote{ningbbo@gmail.com}$~^c$
\\
\\
{\small  $^a$ \it The Institute for Fundamental Study ``The Tah Poe Academia Institute",}
\\
{\small\it Naresuan University, Phitsanulok 65000, Thailand}
\\
\\
{\small $^b$ \it Department of Physics, National Taiwan Normal University, Taipei, 116, Taiwan}
\\
\\
{\small\it $^c$ Center for Theoretical Physics, College of Physical Science and Technology,}
\\
{\small\it Sichuan University, Chengdu, 610064, PR China}
}

\title{\bf Pseudo-topological Quasi-local Energy of Torsion Gravity }

\date{}
\begin{document}
\maketitle

\abstract{ 

   Torsion gravity is a natural extension to Einstein gravity in the presence of the fermion matter sources. In this paper we adopt Wald's covariant method of Noether charge to construct the quasi-local energy of the Einstein-Cartan-fermion system, and find that  its explicit expression is formally independent of the coupling constant between torsion and axial current. This seemingly topological nature is unexpected and is reminiscent of similar nature of quantum Hall effect and topological insulator. However, the coupling dependence does enter when evaluating it on-shell, and thus the topological nature is pseudo. Based on the expression of the quasi-local energy, we evaluate it for a particular solution on the entanglement wedge and find the agreement with the holographic relative entropy obtained before. This shows that the equivalence of these two quantities in the Einstein-Cartan-fermion system. Moreover, the quasi-local energy in this case is not always positive definite so that it provides an example of swampland in torsion gravity.  Based on the covariant Noether charge, we also derive the nonzero fermion effect on Komar angular momentum. The implication of our results to the tests of torsion gravity in the future gravitational wave astronomy is also discussed.

\thispagestyle{empty}
\newpage
\tableofcontents

\setcounter{equation}0

\newpage

 \section{Introduction}

    The general covariance is the key ingredient in formulating the general theory of relativity, which can be thought of as the infinite dimensional local gauge symmetry. However, due to such a huge gauge symmetry, it is then impossible to define any sensible local observable in the context of general relativity, such as the stress tensor associated with the gravitational field.  Instead, some global or quasi-local quantities were proposed, for example, the total mass/energy of the self-gravitating systems. The first quantity is the well-known ADM mass \cite{Arnowitt:1960zzc} and it is useful in formulating the black hole thermodynamics \cite{Bardeen:1973gs,Gibbons:1976ue}. The success was then inspired to find the quasi-local energy \cite{Brown:1992br} covering only a finite domain of the spacetime, which is more close to the classic notion of the local energy density.   There are various ways of deriving quasi-local energy, for a review see \cite{Szabados:2004vb}. Among them the derivation based on Wald's formulation \cite{Wald:1993nt,Iyer:1994ys,Iyer:1995kg,Wald:1999wa,Hollands:2012sf} has the advantage of obtaining a covariant quasi-local energy as Noether charge associated with some time-like Killing vector.

    Recently, the quasi-local energy in the AdS space is proposed to be equivalent to the relative entropy of the dual CFT in the context of Ryu-Tagayanagi's proposal of holographic entanglement entropy \cite{Ryu:2006bv,Ryu:2006ef}. In particular, the positive energy condition of the quasi-local energy is shown to be the same as the positivity of the relative entropy, which can then be used to constrain the swampland of the bulk gravity theory \cite{Lashkari:2015hha,Lashkari:2016idm}. It implies a deep connection between positive energy theorem \cite{Schon:1979rg,Schon:1981vd,Witten:1981mf,Gibbons:1982jg} of gravity theory and the quantum information inequalities of CFTs. It is further shown that the holographic relative entropy for the usual Einstein gravity in AdS space can be constructed as Wald's quasi-local energy.   We then expect that the proof in \cite{Liu:2003bx} for the positive energy condition of quasi-local energy    for flat space can be generalized to AdS space, and similarly the proof in \cite{Abbott:1981ff,Gibbons:1982jg}  of the proof of positive ADM mass in AdS space to the quasi-local energy.
    
     On the other hand, the positive energy theorem for gravity theories other than Einstein gravity and its connection to the positivity of relative entropy in dual CFTs is less explored.   One inspiring example is done by two of us in \cite{Lin:2016fua}, in which we studied the holographic relative entropy of the deformed CFT in the bulk Einstein-Cartan gravity, i.e., torsion gravity \cite{Hehl:1976kj}. We first solve perturbatively the field equations of Einstein-Cartan-fermion system up to second order of Newton constant. Based on this solution, we can evaluate the variations of the modular Hamiltonian and entanglement entropy, and then use these to obtain the relative entropy of the dual CFT.  Interestingly, we find that the resultant relative entropy is not always positive definite so that it implies a swampland in the bulk torsion gravity possibly beyond the reach of weak gravity conjecture \cite{ArkaniHamed:2006dz}.

    In this paper we will explicitly construct the variation of quasi-local energy for Einstein-Cartan-fermion system, and demonstrate the equivalence between holographic relative entropy and quasi-local energy in torsion gravity. To achieve this, we work on Wald's formalism to derive the quasi-local energy not in its original formulation in terms of metric but in terms of vielbein and spin connection. We then evaluate the derived quasi-local energy on the entanglement wedge for a particular perturbative solution and show the agreement with the holographic relative entropy obtained in \cite{Lin:2016fua}.  Besides, our results also show that the quasi-local energy is not always positive, i.e., a swampland, even the theory itself is consistent with the symmetry principle.        
      
    Moreover, the symplectic potential as well as Noether charge are found to be \textit{formally} independent of the torsion-fermion (axial current) coupling constant. Hence, there is no direct torsion contribution to the physical charges such as quasi-local energy or ADM quantities. The torsion infiltrates into the value of physical charges only through the back reacted on-shell solution.  It suggests that the physical charges constructed via Wald's formalism are ``pseudo-topological quantities", i.e., their values are somehow stable against the change of the torsion-fermion coupling. This is a reminiscent of the topological order in quantum Hall systems\cite{wen} or topological insulators \cite{kane} for which the physical quantities are insensitive to some coupling strength and there is a bulk-edge correspondence \cite{spt}.     

   As a natural extension of Einstein gravity, torsion gravity calls for arenas to test its validity, and we think the results obtained and the techniques developed in this paper should be helpful for this purpose. For example, there are more serious attempts of incorporating torsion effect in the cosmological models under scrutiny of CMB physics, see \cite{Cai:2015emx} for a review.    
 Another arena is the gravitational wave astronomy expected by the more coming events similar to the recent LIGO discoveries of gravitational waves emitted from the compact binaries \cite{Abbott:2017vtc, Abbott:2016nmj, TheLIGOScientific:2016src}. Once there are enough events to reduce statistical uncertainties, it should be able to test the validity of Einstein gravity and some modified gravities such as torsion gravity. For example, an analysis of the constraints by the first two LIGO observations on physics beyond Einstein is already put forward in \cite{Yunes:2016jcc}. This calls for more precise theoretical templates of gravitational waves for compact binaries to fitting the observed data. As most of the templates at this stage are done for Einstein gravity, there remains a lack of the high precision templates for the modified gravity theories. Our construction of quasi-local energy, ADM mass and angular momentum can be seen as the first step toward this challenging goal in torsion gravity. For example, one can use these quantities to construct the effective field theory for the coupling between torsion and spin for the post-Newtonian approximation such as done for the coupling between spin and spin connection in Einstein gravity \cite{Steinhoff:2008zr,Steinhoff:2008ji,Steinhoff:2014kwa}. Moreover, these  conserved quantities can also serve as the adiabatic invariants in the framework of effective-one-body approach \cite{Buonanno:1998gg,Damour:2011xga}, which has been used to generate most of waveform templates in Einstein gravity. One more challenging task is to generalize the BSSNOK formulation of numerical gravity \cite{Shibata:1995we, Nakamura:1987zz, Baumgarte:1998te} to a first-order formulation in terms of vielbeins and spin connections because the fermion couples to spin connection not the metric. Our extension of Wald's formalism in using local tetrads will be helpful to this end.

The remaining of the paper is organized as follows.
In the next section, we will briefly review Wald's formalism and torsion gravity. The section \ref{sec 3} contains our main results. We first generalise Wald's formalism to the case with fermion matter and torsion coupling, and then compute the quasi-local energy of the entanglement wedge to compare with the relative entropy. We also discuss the effects of torsion and fermion on some ADM quantities, in particular the extension of Komar angular momentum. In section \ref{sec 4}, we discuss the implication to gravitational wave physics. We then conclude our paper in section \ref{conclusion}. In Appendix \ref{app:xi} we give the details of solving the deformed Killing vector field for evaluating the quasi-local energy.

\section{Wald formalism and torsion gravity} 
Before applying Wald formalism to the Einstein-Cartan-fermion system to obtain our main result in Section \ref{sec 3}, we provide concise reviews on each separately. 

\subsection{Wald formalism for quasi-local energy}\label{Wald-formal}
In this section, we briefly review the quasi-local energy defined via the covariant phase space formalism put forward by Wald and his collaborators \cite{Wald:1999wa,Iyer:1994ys}. The basic idea is to construct a covariant Noether current and charge associated with a time-like vector field inside a space-like subregion, and then relate this Noether charge to the quasi-local energy defined for this subregion.

Let us denote the subregion of the Cauchy surface by $\S$ and denote all the dynamical fields (including metric\footnote{The original formulation in the literature is developed for metric gravity, however, we will see in Section \ref{sec:derQE} that Wald formalism can be generalized to theories formulated with vielbein.} 
and matter fields) collectively as $\f$. 
In the following, a boldface letter denotes a differential form in the space-time, 
for example, the Lagrangian is written as a 4-form $\bL$. 
Generically, the variation of a covariant Lagrangian is written as 
\be\label{symplectic}
\d\bL = \bE \d\f + d\bTh(\f,\d\f), 
\ee
where $\bE = 0$ are the field equations and the surface term $\bTh(\f,\d\f)$, called the symplectic potential 3-form, is constructed covariantly and locally in terms of $\f$ and $\d\f$. 
The following anti-symmetrised variations of $\bTh$ gives rise to the symplectic current 3-form
\be
\bo(\f,\d_1\f,\d_2\f) = \d_1\bTh(\f,\d_2\f) - \d_2\bTh(\f,\d_1\f). 
\ee
Note that the symplectic current is a bilinear functional of $\d\f_1$ and $\d\f_2$, and its volume integral is simply the symplectic form. 

    Given an arbitrary vector field $\xi$ we can formally associate with it a Hamiltonian $H_\xi$, with its variation satisfying 
\be
\d H_\xi = \int_\S \bo(\f,\d\f,\cL_\xi\f), 
\ee
where $\cL_\xi\f$ is the Lie derivative of $\f$ along the vector field $\xi$. If $\xi$ is a time-like vector field, it is natural to interpret $\d H_\xi$ as the perturbation of the quasi-local energy contained in the subregion $\S$ \cite{Lashkari:2016idm} \footnote{As commented in \cite{Lashkari:2016idm}, this is a natural generalization of Hamiltonian for the particle Lagrangian. In \cite{Hollands:2012sf,Lashkari:2015hha} $\d H_\xi$ is called canonical energy for the second order perturbation, however, this could be confused with the linearized ADM mass for which it was also called canonical energy in \cite{Wald:1993nt,Iyer:1994ys}. Thus, we will simply call $H_\xi$ quasi-local energy to avoid confusion.}. 
Moreover, the existence of the full Hamiltonian $H_\xi$ requires the following integrability condition 
\be \label{intcond}
0 = (\d_1\d_2 - \d_2\d_1) H_\xi 
 = -\int_{\pa\S} \xi\cdot \bo(\f,\d_1\f,\d_2\f). 
\ee

    On the other hand, we can associate with $\xi$ a Noether current 3-form defined by
\be
\bJ_\xi = \bTh(\f,\cL_\xi\f) - \xi\cdot\bL. 
\ee
It is straightforward to show that Noether current 3-form is closed on-shell, i.e., $d\bJ_\xi=-\bE \cL_\xi\f$, so that it can be written as \cite{Iyer:1995kg}
\be
\bJ_\xi = d\bQ_\xi + \xi^\m \bC_\m, 
\ee
where $\bC_\m$ vanishes on-shell. The space-time 2-form $\bQ_\xi$ is the Noether charge.    
A useful identity relates the symplectic current to the variation of the Noether current is given by 
\be
\bo(\f,\d\f,\cL_\xi\f) = \d \bJ_\xi - d\lrbrk{\xi\cdot\bTh(\f,\d\f)}, 
\ee
with $\f$ assumed to be on-shell.   Using this, the symplectic current can be written as 
\be
\bo(\f,\d\f,\cL_\xi\f) = d \lrbrk{\d\bQ_\xi - \xi\cdot\bTh} + \xi^\m \d\bC_\m. 
\ee

Hence, if we restrict to variations $\d\f$ that satisfy the field equations (so that $\d\bC_\m = 0$), we obtain the expression for the variation of the quasi-local enery 
\be  \label{qEsec2}
\d H_\xi = \int_{\pa\S} \lrbrk{\d\bQ_\xi - \xi\cdot\bTh}. 
\ee
Notice that this is in a form of a surface integral. In general, $\pa\S$ contains 2 parts: one at asymptotic infinity denoted by $B$, and the inner boundary into the bulk denoted by $\Bt$.  Moreover, the form of \eq{qEsec2} suggests that the integrability condition \eq{intcond} should be equivalent to the existence of some $\bK$ such that
\be  \label{defbK}
\d (\xi \cdot \bK) =\xi \cdot \bTh \qquad  \mbox{on} \; \pa\S\;.
\ee
If so, we then have the full quasi-local energy
\be\label{Hxi}
H_\xi = \int_{\pa\S} \lrbrk{\bQ_\xi - \xi\cdot\bK}\;, 
\ee
and the difference of quasi-local energy between two geometries is given by 
\be  \label{DeltaqE}
\Delta H_\xi 
= \Delta \int_{\pa\S} \lrbrk{\bQ_\xi - \xi\cdot\bK}\;. 
\ee

   In the case that $\xi$ is a Killing vector field, i.e., $\cL_\xi \f=0$ so that $\bo(\f,\d\f,\cL_\xi\f=0)=0$ hence $\d H_\xi=0$. Then, \eq{qEsec2} can be written as the form of first law
\be \label{first law}
\int_{B} \lrbrk{\d\bQ_\xi - \xi\cdot\bTh}=\int_{\Bt} \lrbrk{\d\bQ_\xi - \xi\cdot\bTh}\qquad  \Leftrightarrow \qquad \d \cE={\kappa_s \over 8\pi G_N} \d \cA \;.
\ee
The LHS is related to variation of canonical energy $\d \cE$ such as ADM mass or the modular energy, and the RHS is related to the variation of the  area $\d \cA$ of the inner boundary $\Bt$. To make the first law manifest we should impose the following boundary conditions on $\xi$:
\bea 
\xi \big|_{B} &=& \zeta,  \label{xionB} \\ \label{binormal-c}
\tilde{\nb}^{[\m}\xi^{\n]} \big|_{\Bt} &=& \kappa_s n^{\m\n}, \\ 
\xi \big|_{\Bt}  &=& 0, \label{xibcsec2}
\eea
where $\tilde{\nb}^\n$ is the Riemannian covariant derivative, $\zeta$ is the asymptotic time-like Killing vector field, $\kappa_s$ is the surface gravity for the inner boundary $\Bt$ and $n^{\m\n} := n_\1^\m n_\2^\n - n_\2^\m n_\1^\n$ is the unit binormal vector.  
For the black hole geometry, $B$ is the full asymptotic boundary and $\Bt$ is the black hole horizon. Thus, \eq{first law} will yield the first law of black hole thermodynamics. On the other hand, if we consider the asymptotic AdS space, then $\S$ is the entanglement wedge bounded by the asymptotic boundary disk $B$ and the Ryu-Takayanagi minimal surface $\Bt$ whose surface gravity $\kappa_s$ is set to $2\pi$. Thus, \eq{first law} yields the first law of the entanglement thermodynamics for the dual CFT. 
  
     In this paper we will consider the case that $\xi$ is not the Killing vector field for the background solution, i.e., $\cL_\xi \f \ne 0$  so that $\d H_\xi$ can be treated as the quasi-local energy for $\phi$.   Despite that, to preserve the asymptotic Killing symmetry we will impose the additional boundary condition
\be\label{add-bc}
\cL_\xi \f \big|_{B}=0\;.
\ee
If $\xi$ is a time-like vector field, this requires $\f$ to be asymptotically stationary.
A specific example to be considered below is the holographic relative entropy, which is dual to the quasi-local energy \eq{qEsec2} for the entanglement wedge in the AdS space for torsion gravity.   This quasi-local energy is for the second order stationary solution so that the background metric is the metric up to the first order, and hence $\cL_\xi \f = 0$ does not hold (though \eq{add-bc} still holds) to yield nonzero $\d H_\xi$.

\subsection{Torsion gravity}

   In this subsection we briefly review the torsion gravity. We start with the Lagrangian for Einstein-Cartan-fermion system:
\be\label{totalL}
\bL
= \ove{2\k^2} \bL_R + \bL_M, 
\ee
where $\kappa^2:=8\pi G_N$ and the Einstein-Cartan part $\bL_R$ and the fermion part $\bL_M$ of the Lagrangian are respectively given by 
\bea\label{LR-1}
\bL_R
&= &{\color{black}-} 
	e^\m_a e^\n_b R_{\m\n}{}^{ab} \bep - 2\L \bep = (R-2\L) \bep, \\
\bL_M \label{LM-1}
&= & -\frac{1}{2} 
	\lrsbrk{ \bar\psi \g^\m\nb_\m\psi - (\nb_\m\bar\psi)\g^\m\psi + 2m\bar\psi\psi } \bep
\eea
with the vielbein $e^\m_a$ and the volume element $\bep := \sqrt{-g}d^4x$.

  The covariant derivative for the Dirac fermion field $\psi$ and the curvature tensor used in \eq{LR-1} and \eq{LM-1} are formally defined as usual in terms of the spin connection $\om_\m{}^{a}{}_b$, e.g., the Riemann tensor
\be  \label{cpntRmnab}
R_{\m\n ab}
= - \pa_\m \om_{\n ab} + \pa_\n \om_{\m ab} 
	- \om_{\m ac}\om_\n{}^c{}_b +  \om_{\n ac}\om_\m{}^c{}_b. 
\ee
However, the spin connection now contains the torsion part. Explicitly it can be divided into the following:
\be\label{spinconnection-torsion}
\om_{\m\n\r}:= \om_{\m ab}e^a_\n e^b_\r = \tilde{\om}_{\m\n\r}(e) + K_{\m\n\r}, 
\ee
where the Riemannian part of spin connection is given by 
\be
\tilde{\om}_{\m}{}^{ab}(e)= 2e^{\n[a}\pa_{[\m}e^{b]}_{\n]} - e^{\n[a}e^{b]\s}e_{\m c}\pa_{\n}e^c_\s,
\ee
and the remaining part is the contorsion tensor $K_{\m\n\r}$ which is related to the torsion tensor 
$S_{\m\n}{}^{\r}:={1\over 2}(\Gamma_{\m\n}^{\r}-\Gamma_{\n\m}^{\r})$ with $\Gamma_{\m\n}^{\r}$ the affine connection by the following:
\be
K_{\m\n\r} = - (S_{\m\n\r} - S_{\n\r\m} + S_{\r\m\n}). 
\ee
In the following, we will work on the formalism developed by \cite{Freedman:2012zz}, in which the the vielbein $e^a_\m$ and the torsion tensor $S_{\m\n}{}^{\r}$ are considered as independent fields. 
 
  Moreover, we can introduce the non-minimal coupling between fermion and torsion in the following way:
\be  \label{nbsnbpsi}
 \nb_\m\psi \rar \nbs_\m\psi 
 := \pa_\m\psi + \ove{4}\tilde\omega_\m{}^{ab}\g_{ab}\psi + \frac{\eta_t}{4}K_{\m\n\r}\g^{\n\r}\psi \;,
 \ee
and similarly for $\nbs_\m\bar\psi:= \nb_\m\bar\psi - \frac{\eta_t-1}{4}K_{\m\n\r}\bar\psi\g^{\n\r}$. It is called minimal coupling when $\eta_t=1$.  This amounts to adding in the following interaction term:
\be
-{1\over 4} (\eta_t-1) \sqrt{-g} \,{\bar \psi} \g^{[\m} \g^{\n} \g^{\l]} \psi K_{\m\n\l}.
\ee
In \cite{Lin:2016fua} it was shown that there is nontrivial constraint on $\eta_t$ from the positivity of holographic relative entropy. In this paper we will show that the same constraint arises from the positivity of quasi-local energy over the entanglement wedge.

 By the variation principle we obtain the field equations for the action $\bL$:
\be  \label{EOMS2nd}
S^{\m\n\r} =\eta_t  \frac{\k^2}{4} \bar\psi\g^{\m\n\r}\psi, 
\ee
\be  \label{EOMpsi2nd}
\overset{*}{\nb}_\m\bar\psi\g^\m - m\bar\psi = 0, \qquad 
\g^\m\overset{*}{\nb}_\m\psi + m\psi = 0, 					
\ee
\be  \label{EOMe2nd}
G_{\m\n} + \L g_{\m\n} = \k^2 \lrbrk{ \bar\Sigma_{(\m\n)} + \eta_t \bar\Sigma_{[\m\n]} }\;.
\ee
where $\bar\Sigma_{(\m\n)}$  and $\bar\Sigma_{[\m\n]}$ are symmetric and ant-symmetric parts of $\bar\Sigma_{(\m\n)}$, which is defined by 
\be
 \bar\Sigma_{\m\n}:= \ove{2}\lrsbrk{\bar\psi\g_\n\nbs_\m\psi- (\nbs_\m\bar\psi)\g_\n\psi}\;.
\ee

  To solve the field equations, one can split the Einstein tensor into the Riemannian part as well as the non-Riemannian part. In \cite{Lin:2016fua} we have done this way to obtain the second order perturbative solution in asymptotically AdS space to evaluate the holographic relative entropy. The solution is summarized as follows. 
  
  To set up the notation, the AdS metric is put in the Poincare coordinate
\be\label{AdSmetric}
ds^2 = \frac{\ell^2 }{z^2} \left( - dt^2 +  dx^2 + dy^2 + dz^2 \right)\;.
\ee
We will expand the solution in terms of dimensionless Newton constant
\be
\sk:={\kappa^2 z_L^2 \over \ell^4}
\ee
where $z_L$ is an IR cutoff, however all the physical quantities such as quasi-local energy are independent of it.

  Then, the fermion solution up to first order in $\sk$ is as follows:
\be\label{fmode}
\psi  =  \left( \begin{array}{c} ({z\over \ell^2})^{3/2 \,+\,  m \ell} \,a_+ \\  ({z\over \ell^2})^{3/2 \,-\, m \ell} \,a_- \end{array}\right)+  \frac{\sk}{3} \left( \begin{array}{c} \,~\Delta_+ \,({z\over \ell^2})^{9/2 \,+\, m \ell} \,a_+ \\
-\Delta_- \, ({z\over \ell^2})^{9/2 \,-\, m \ell} \,a_-\end{array}\right)\,,
\ee
where 
\be
\Delta_{\pm} =   {1\over 4} \Big(3 \eta_t^2 \,+\, 2 \mu_0 m^2 \ell^2 \pm (3 \mu_0 - 2) m \ell \Big) \ell^5 \a \b z_L^{-2}\;,
\ee
with $\mu_0$ the integration constant while solving the first order metric. 
Besides, the $a_{\pm}$ are integration constant 2-spinors, and we will choose $a_{\pm}$ the following without loss of generality:
\be \label{IntC}
 a_+ = \{0,\, \a\}^{\mathrm{T}}, \qquad a_- = \{i \b,\, 0\}^{\mathrm{T}}.
\ee

Due to the presence of the fermion solution \eq{fmode}, the AdS metric \eq{AdSmetric} is then backreacted into 
\be\label{backreactg}
ds^2 = \frac{\ell^2 }{z^2} \left( - F(z) \,dt^2 + H(z)\, dt \,dx +  dx^2 + dy^2 + G(z)\, dz^2 \right) \,, 
\ee
with 
\bea 
F(z) &=& 1 + \sk\, \left( {2 \over 3} - \mu_0  \right) m  \a \b \, {z^3 \over z_L^2} + \sk^2\,  \frac{(2 - 3 \mu_0)\, m^2 \a^2 \b^2 }{18}\, {z^6 \over z_L^4} \,, \nn\\
G(z) &=& 1+ \sk\, \mu_0 m \a \b \, {z^3 \over z_L^2} + \sk^2\, \frac{(\eta_t^2 \ell^{-2} + 4 \mu_0^2 m^2 )  \a^2 \b^2}{4}\, {z^6 \over z_L^4} \,, \label{gzz} \\
H(z) &=& \sk^2\, \frac{(2 - 3 \mu_0)\, m \ell^{-1} \a^2 \b^2}{18} \, {z^6 \over z_L^4} \,. \nn
\eea

        
\section{Topological quasi-local energy of torsion gravity}\label{sec 3}

  In this section, we present our results starting from the derivation of quasi-local energy for Einstein-Cartan-fermion system in section \ref{sec:derQE}.  Then, the explicit computation of it on the entanglement wedge for the solution \eq{backreactg} and the comparison with relative entropy are presented in section \ref{sec:eva}. Finally the torsion and fermion effects in ADM mass and angular momentum are studied in section \ref{sec:ADM}. 

Wald's covariant phase space formulation as reviewed in sec. \ref{Wald-formal} is very powerful and has been widely applied to black hole thermodynamics and study of holographic entanglement entropy. However, the discussions in the literatures have been restricted in metric gravity. In section \ref{sec:derQE}, we show how the formalism can be extended to including fermions with torsion coupling. Readers who just want to plowing ahead to the physical implications of the result may skip this technical section \ref{sec:derQE}, although these techniques could be useful in the context of gravitational wave physics of torsion gravity. Subsequently, we show that the quasi-local energy of torsion gravity is not always positive definite but instead leads to a bound constraining the physical parameters of the theory. Remarkably, this means an innocent looking theory which passed all the symmetry constraints might be actually pathological. The same bound has been obtained in  \cite{Lin:2016fua} in a different context by the holographic computation of the relative entropy. Our results provide a nontrivial example of swampland beyond the reach of the grand symmetry principle and possibly the weak gravity conjecture \cite{ArkaniHamed:2006dz}.

In the end, we discuss the fermion and torsion effects on the ADM mass and angular momentum. We found that the angular momentum is extended by the axial current in the asymptotically flat space. The physical gauge invariant quantities such as global charges are crucial in many aspects of gravity. For example, we expect this extension term plays an important role in the canonical analysis of torsion gravity and hence deforms the post-Newtonian expansion of the gravitational waveform as well as the adiabatic invariants in the framework of effective-one-body approach \cite{Buonanno:1998gg}. This will be further explored in the future works.

\subsection{Derivation of quasi-local energy for torsion gravity}  \label{sec:derQE}

  To proceed with Wald's formalism, we need to variate the action carefully with all the surface terms arising from integration by parts retained.   In this subsection, we  first present the essential steps of extracting the symplectic potential by variating the action of Einstein-Cartan-fermion system with respect to the independent fields $e^a_\m$, $S_{\m\n}{}^a$ and $\bar\psi$, $\psi$. 
To the best of our knowledge, this is the first discussion on the quasi-local energy of fermionic fields coupled to torsion in the literatures, and thus we pull down the details which could be useful for other explorations. Then, based on the result of symplectic potential we derive the associate Noether charge and the variation of the quasi-local energy. 

\subsubsection{Summary of the results}
   Although the procedure seems straightforward, it is in fact quite tedious.  Before sketching the detailed derivation, we first write down the result:  the symplectic potential defined in \eq{symplectic} for the Einstein-Cartan-fermion system turns out to be 
\be  \label{bThEC}
\begin{split}
\bTh(\phi,\d \phi)
&= \ove{3!}\ove{2\k^2}\bep_{\m\r_1\r_2\r_3}
	\bigg(g^{\m\a}g^{\b\g} 
			(\tilde\nb_\b\d g_{\a\g} - \tilde\nb_\a\d g_{\b\g})   \\
&\qquad\qquad\qquad
		  - {\k^2 \over 2} \bar\psi\g^{\a\g\m}\psi \d e^a_\a\, e_{\g a}
		  + \k^2 (\d\bar\psi\g^\m\psi - \bar\psi\g^\m\d\psi) \bigg)  dx^{\r_1\r_2\r_3},
\end{split}
\ee
where $dx^{\r_1\r_2\r_3}$ is a shorthand for the wedge product $dx^{\r_1}\w dx^{\r_2} \w dx^{\r_3}$. Recall that $\tilde{\nb}^\n$ is the Riemannian covariant derivative so that \eq{bThEC} is reduced to the symplectic potential for pure Einstein gravity once the  fermion field is put to zero. Moreover, the last two terms are sub-leading order in $\k^2$ compared to the first term.   

 Then, we will use this symplectic potential to obtain the Noether charge associate with some vector field $\xi$. The result is
\be  \label{bQEC}
\bQ 
= \ove{2!} \frac{-1}{2\k^2} \bep_{\a\b\r_1\r_2} 
		\lrbrk{\tilde{\nb}^\a \xi^\b 
			  + \k^2\ove{4}\bar\psi\g^{\a\b\g}\psi\, \xi_\g } dx^{\r_1\r_2}. 
\ee
Use the above we can further obtain the quasi-local energy or its variation by \eq{Hxi} and \eq{qEsec2}.  Note that we have used on-shell relation \eq{EOMS2nd} in arriving \eq{bThEC} and \eq{bQEC} by replacing torsion with fermion bilinear and at the same time cancelling the $\eta_t$ dependence. 

Before starting the derivation, we remark that the symplectic potential \eq{bThEC} and the Noether charge \eq{bQEC} both are formally independent of the torsion coupling $\eta_t$, so is the quasi-local energy. This is intriguing because it implies the quasi-local energy, which is a physical quantity, is \textit{formally} independent of the torsion coupling. The only way that torsion takes effect is through the back reacted geometry. 
The analogy that shares this feature is the topological order observed in the quantum Hall systems or topological insulators. This analogy is just formal as the topological order is known to be due to the nontrivial patterns of many-body entanglement. On the other hand, the quasi-local is a classical quantity of gravity theory. 

Also notice that $\eta_t$ cannot be absorbed by field redefinitions. This is actually expected because the physical quantities, such as relative entropy that is calculated in \cite{Lin:2016fua} and the quasi-local energy that will be computed later depend on $\eta_t$ explicitly. The appearance of the $\eta_t$ dependence in the value of quasi-local energy, however, comes from the $\eta_t$ dependence of the backreacted on-shell solution though the formal expression of the quasi-local energy is independent of $\eta_t$. 

\subsubsection{Variation of Einstein-Cartan action}
   We first consider the variation of the Einstein-Cartan action.  The Ricci curvature can be expressed in terms of the covariant derivative of spin connection. We can variate the Ricci curvature with respect to vielbein and spin connection, and then relate the variation of the spin connection to the ones of vielbein and torsion by the following:
\be  \label{varC1SO}
e_{a\n}e_{b\r}\d\om_\m{}^{ab}
= \triangle^{\t\s\l}_{\r\m\n} \nb_{[\t} \d e^a_{\s]} e_{a\l} 
	+ \triangle^{\t\s\l}_{\r\m\n} S_{\t\s}{}^\eta \d e^a_{\eta} e_{a\l} 
	- \triangle^{\t\s\l}_{\n\r\m} e_{a\t} \d S_{\s\l}{}^a
\ee   
where
\be
\triangle^{\t\s\l}_{\r\m\n} 
:= \d^\t_\r \d^\s_\m \d^\l_\n - \d^\t_\n \d^\s_\r \d^\l_\m 
	+  \d^\t_\m \d^\s_\n \d^\l_\r. 
\ee

  Using the above relations and the fact $\d e^\m_a R_\m{}^a = -R_a{}^\m \d e^a_\m 
$, we can arrange the variation of $\bL_R$ into the following:
\be \label{debLR2nd}
\begin{split}
\d \ove{2\k^2}\bL_R 
&= 
	 - \ove{\k^2} (G_a{}^\m+\L e^\m_a) \bep \d e^a_\m
	 + d \lrbrk{ \ove{3!}\ove{\k^2}\bep_{\m\b\g\d} 
	 		e^{a\m}e^{b\n}\d\om_{\n ab} dx^{\b\g\d} }  \\
&\qquad -\frac{2}{\k^2} \bep\, S^\m{}_\r{}^\r g^{\eta\n}
			\lrbrk{ \triangle^{\t\s\l}_{\eta\n\m} \nb_{[\t}\d e^a_{\s]}e_{a\l}
				   + \triangle^{\t\s\l}_{\eta\n\m} S_{\t\s}{}^\th \d e^a_{\th} e_{a\l} } \\
&\qquad +\frac{2}{\k^2} \bep 
			\lrbrk{ S^\n{}_\s{}^\s g^{\r\m} 
				  - S^\r{}_\s{}^\s g^{\m\n} 
				  + S^\m{}_\s{}^\s g^{\n\r} } e_{a\n} \d S_{\r\m}{}^a    \\
&\qquad + \frac{1}{\k^2}S^{\m\n\r} 
		\lrbrk{ \triangle^{\t\s\l}_{\n\r\m} \nb_{[\t}\d e^a_{\s]} e_{a\l}
			   + \triangle^{\t\s\l}_{\n\r\m} S_{\t\s}{}^\eta \d e^a_{\eta} e_{a\l} } \bep   \\
&\qquad - \frac{2}{\k^2} \bep S^{(\n|\r|\m)} e_{a\n} \d S_{\r\m}{}^a 
		- \ove{\k^2} \bep S^{\m\n\r} e_{a\n} \d S_{\r\m}{}^a. 
\end{split}
\ee
In the process, one should keep track of all surface terms arising from integration by parts. To this end, it is useful to use the modified divergence operator 
\be
\hat{\nb}_\mu := \nb_\mu + 2S_{\m\n}{}^\n
\ee
which satisfies 
\be
\hat{\nb}_\mu v^\mu = \pa_\mu v^\mu 
\ee
but not Leibniz rule. 
The $v^\m$ in the above formula is a vector density.  Therefore, the surface term in \eq{debLR2nd} in fact comes from the following
\be   \label{surLR2nd}
\hat{\nb}_\m \lrbrk{ \frac{e}{\k^2} e^{a\m} e^{b\n}\d\om_{\n ab} }d^4x
= \pa_\m \lrbrk{ \frac{e}{\k^2} e^{a\m} e^{b\n}\d\om_{\n ab} }d^4x = d \lrbrk{ \ove{3!}\ove{\k^2}\bep_{\m\b\g\d} e^{a\m}e^{b\n}\d\om_{\n ab} dx^{\b\g\d} }. 
\ee
The other terms in \eq{debLR2nd} will be combined with non-surface terms in $\d\bL_M$ into field equations.

\subsubsection{Variation of fermion action}
   Now, let us consider the variation of the fermion action. This is quite similar to the usual variation of Dirac fermion action in curved space except that now the fermion also couples to torsion. However, it is straightforward to see that the torsion does not contribute to surface term through this variation.   After some tedious calculations, we obtain the result as follows:  
\be   \label{dbLM2nd}
\begin{split}
\d\bL_M 
&= \d\lrbrk{ \bep \cL_M} 
= \ove{2}g^{\m\n} \d(e^a_\m\eta_{ab}e^b_\n) \bep \cL_M 
	+ \bep \d\cL_M  \\
&= \bep\cL_M e^\m_a \d e^a_\m  
+d\lrsbrk{ \ove{3!}\ove{2}\bep_{\m\b\g\d} 
			 \lrbrk{\d\bar\psi\g^\m\psi - \bar\psi\g^\m\d\psi} dx^{\b\g\d} } \\
&\quad + \bep \lrsbrk{ \lrbrk{\nbs_\a\bar\psi\g^\a - m\bar\psi 
							+  S_{\a\b}{}^\b \bar\psi\g^\a}\d\psi  
					 - \d\bar\psi\lrbrk{\g^\a\nbs_\a\psi + m\psi 
					 				{\color{black}\,+\,}  S_{\a\b}{}^\b \g^\a\psi} }  \\
&\quad - \ove{2}\bep \, \bar\psi\g^a \overset{\lrar}{\nbs}_\m \psi \d e^\m_a
	- \ove{4}\bar\psi\g^{\m\n\r}\psi
		 \lrbrk{ \nb_\r \d e^a_\m e_{a\n}
				+ S_{\r\m}{}^\s \d e^a_\s e_{a\n}
				- \eta_t e_{a\n}\d S_{\r\m}{}^a } \bep \\
&\quad + \frac{\eta_t-1}{4} \bep (\bar\psi \g^{b\n\r}\psi) S_{\m\n\r} \d e^\m_b \;,							
\end{split}
\ee
where $\bL_M := \cL_M \bep$. 
In arriving the above, we have used the relation \eq{varC1SO}.

   The $\d \psi$ and $\d\bar\psi$ terms in \eq{dbLM2nd} give the field equation \eq{EOMpsi2nd}. On the other hand, by combining the non-surface terms associated with $\d S_{\r\m}{}^a$ in both \eq{debLR2nd} and \eq{dbLM2nd}, we obtain the field equation for torsion field:
\be  
\lrsbrk{ -2S^{(\n|\r|\m)} - S^{\m\n\r} 
		+ 2S^\n{}_\s{}^\s g^{\r\m} - 2S^\r{}_\s{}^\s g^{\m\n} + 2S^\m{}_\s{}^\s g^{\n\r} 
		+ \eta_t {\k^2 \over 4}\bar\psi\g^{\m\n\r}\psi}
\bep \, e_{a\n} = 0.  
\ee
It is straightforward to solve it and arrive at \eq{EOMS2nd}.  

   Next, we combine the terms involving $\nb_\r \d e^a_\m$ in \eq{debLR2nd} and \eq{dbLM2nd} with the help of \eq{EOMS2nd}, and then integrate by part to obtain the additional surface term. Explicitly, the combined result gives  
\be\label{nbeam}
\begin{split}
& \frac{\eta_t-1}{4} (\bar\psi\g^{\m\n\r}\psi)  e_{a\n} \nb_\r\d e^a_\m \; \bep \\
=&\,d \lrbrk{ \ove{3!}\bep_{\r\a_2\a_3\a_4} \frac{\eta_t-1}{4} 
			(\bar\psi\g^{\m\n\r}\psi) e_{a\n}\d e^a_\m dx^{\a_2\a_3\a_4} } - (\eta_t-1) \Sigma^{[\m\n]} e_{a\n} \d e^a_\m \,\bep, 
\end{split}
\ee
where 
\be
\Sigma_{\m\n}:= \ove{2}\lrsbrk{\bar\psi\g_\n\nb_\m\psi- (\nb_\m\bar\psi)\g_\n\psi}\;.
\ee
Combining the last term on the RHS of \eq{nbeam} with the other terms involving $ \d e^a_\m$ in \eq{debLR2nd} and \eq{dbLM2nd}, we arrive at the field equation \eq{EOMe2nd} after using the on-shell relation 
\be
\Sigma_{[\m\n]} = \bar\Sigma_{[\m\n]}
\ee
upon the use of \eq{EOMS2nd}.

\subsubsection{Combining into Symplectic potential}
   Finally, collecting the surface terms in \eq{debLR2nd}, \eq{dbLM2nd} and \eq{nbeam} we obtain the symplectic potential:  
\bea  \label{bTh2ndtmp}
\bTh 
& = & \ove{3!}\ove{\k^2}\bep_{\m\b\g\d} e^{a\m}e^{b\n}\d\om_{\n ab} dx^{\b\g\d} 
	+ \ove{3!}\ove{2}\bep_{\m\b\g\d}
			 \lrbrk{\d\bar\psi\g^\m\psi - \bar\psi\g^\m\d\psi} dx^{\b\g\d} \nn \\
	&& + \frac{\eta_t-1}{4} \ove{3!}\bep_{\m\b\g\d}
			(\bar\psi\g^{\m\n\r}\psi) e_{a\r}\d e^a_\n dx^{\b\g\d}\; .
\eea
Note that this is not yet the final form of \eq{bThEC}.  To achieve this, we need to use the on-shell relation \eq{EOMS2nd} and the relation \eq{varC1SO} which can be further simplified, by the anti-symmetry of torsion tensor $S^{\a\b\g} = S^{[\a\b\g]}$, into
\be
\begin{split}
e^{a\m}e^{b\n}\d\om_{\n ab}
&= g^{\a\m}g^{\b\n} \triangle^{\t\s\l}_{\b\n\a} \nb_{[\t}\d e^a_{\s]} e_{a\l} \;.
\end{split}
\ee
Then, \eq{bTh2ndtmp} will be turned into \eq{bThEC} by throwing away an exact form 
\be
d \lrbrk{ \ove{4\k^2}\bep_{\m\b\r_1\r_2} 
				g^{\m\a}g^{\b\n}\d e^a_{[\a}e_{\n]a} dx^{\r_1\r_2} }. 
\ee
This ambiguity is allowed as can be seen from the defining equation \eq{symplectic} of symplectic potential because \eq{symplectic} still holds under
\be
\bTh \longrightarrow \bTh + \d {\bf \mu} +d{\bf Y}(\phi, \d \phi)
\ee
as elaborated more in \cite{Iyer:1994ys}. Note that ${\bf \mu}$ is due to the shift of $\bL$: $\bL\longrightarrow \bL+d{\bf \mu}$.

\bigskip

\subsubsection{Obtaining the Noether charge}
We will now derive the Noether current and hence the Noether charge associated with vector field $\xi$, 
and finally the explicit expression of the quasi-local energy for the minimally coupled 
Einstein-Cartan-fermion system. 

  According to the prescription of \cite{Wald:1993nt,Iyer:1994ys},  the Noether current 3-form is given by 
\be\label{NoetherJ}
\bJ_\xi = \bTh(\f,\cL_\xi\f) - \xi\cdot\bL. 
\ee
Our goal is to extract the Noether charge by rewriting $\bJ_\xi$ into the following
\be\label{formalJQ}
\bJ_{\xi}=d \bQ_\xi + (\text{on-shell})
\ee
where the terms in $(\text{on-shell})$ vanish when imposing on-shell condition.

To explicitly carry out the evaluation,  we need the Lie derivatives of the vielbein and fermion field, i.e.,  
\bea  \label{Liepsie-1}
\cL_\xi e^a_\a
&=& \nb_\a\xi^\b \, e^a_\b - \xi^\b \om_\b{}^a{}_c e^c_\a - 2S_{\a\b}{}^a \xi^\b, \\
\cL_\xi\psi 
&=& \xi^\m \pa_\mu\psi  =  \xi^\m \lrbrk{\nbs_\m\bar\psi 
			  - \ove{4}\omega_\m{}^{ab} \g_{ab}\psi
			  - \frac{\eta_t-1}{4}K_{\m\n\r}\g^{\n\r}\psi} .    \label{Liepsie-2}
\eea
Here we adopt the convention in \cite{Freedman:2012zz} for the Lie derivative of the fermion field by treating it like a scalar.  Besides, the following identities are useful in the process of derivation:
\bea \label{identities-1}
\lrsbrk{\nb_\a , \nb_\b} v^\g  &=& R_{\a\b\m}{}^\g v^\m - 2S_{\a\b}{}^\n \nb_\n v^\g, 
\\ \label{identities-2}
\lrbrk{\cL_\xi g}_{\m\n}
&=& 2\nb_{(\m}\xi_{\n)} + 4\xi^\g S_{\g(\m}{}^\b g_{\n)\b} \approx  2\nb_{(\m}\xi_{\n)},    \\ \label{identities-3}
\nb_\g S^{\a\b\g}
&\approx &  - \eta_t \frac{\k^2}{2} \lrbrk{ \nb^{[\a}\bar\psi \g^{\b]}\psi + \bar\psi \g^{[\a}\nb^{\b]}\psi }, 
\\ \label{identities-4}
\xi\cdot\bL_M  &\approx& 0, 
\eea
where $\approx$ denotes the weak equality that holds on-shell.

   We break down the derivation into steps.  We first deal with the first term of \eq{bThEC}, called $\bTh^{(1)}(\phi, \cL_\xi \phi)$. It yields 
\be
\begin{split}
2\k^2\bTh^{(1)}
&=	\ove{3!}\bep_{\m\r_1\r_2\r_3} g^{\m\a}g^{\b\g} 
		\lrbrk{ \nb_\b\nb_\a\xi_\g + \nb_\b\nb_\g\xi_\a 
		 	  - \nb_\a\nb_\b\xi_\g - \nb_\a\nb_\g\xi_\b } dx^{\r_1\r_2\r_3}   \\
&= \ove{3!} \bep_{\m\r_1\r_2\r_3}
	\lrsbrk{ (2R^\m{}_\s\xi^\s + 4S^{\m\b\s}\nb_\s\xi_\b) + 2\nb_\b\nb^{[\b}\xi^{\m]} } 
		 dx^{\r_1\r_2\r_3}  \;.
\end{split}
\ee
In arriving the above, the identities \eq{identities-1} and \eq{identities-2} are used. Moreover, the last term of the last line can be used to make up a total derivative term:
\be   \label{bJRcorrec}
\begin{split}
\ove{3!}\bep_{\m\r_1\r_2\r_3} 2\nb_\b \nb^{[\b}\xi^{\m]} dx^{\r_1\r_2\r_3  }
&= \ove{2!}\nb_{\r_1} \lrbrk{\bep_{\m\b\r_2\r_3} \nb^{[\b}\xi^{\m]} }
		dx^{\r_1\r_2\r_3}   \\
&=
	d \lrbrk{\ove{2!}\bep_{\m\b\r_2\r_3} \nb^\b\xi^\m dx^{\r_2\r_3}}
	+ \ove{3}S^{\b\m\s} \bep_{\s\r_1\r_2\r_3} 
		 \nb_\b\xi_\m dx^{\r_1\r_2\r_3}.
\end{split}
\ee
We have used total antisymmetry of the torsion tensor $S^{\a\b\g} \approx S^{[\a\b\g]}$ to arrive the last equality. 

    Now, comes the second term of \eq{bThEC}, denoted by $\bTh^{(2)}(\phi, \cL_\xi \phi)$, which by using \eq{Liepsie-1} can be further simplified as follows:
\be
\begin{split}
\bTh^{(2)}
 &= {\color{black}\,-\,}  \ove{4!} \bar\psi\g^{\a\n\s}\psi  \; \bep_{\s\r_1\r_2\r_3} 
	\lrbrk{ \nb_{\a}\xi_{\n} 
		   {\color{black}\,-\, \xi^\b\omega_{\b\n\a}}
		   - 2S_{\a\b\n} \xi^\b 
		  } dx^{\r_1\r_2\r_3}. 
\end{split}
\ee
 Finally, we now deal with the last term of \eq{bThEC}, called $\bTh^{(3)}(\phi, \cL_\xi \phi)$.    Using  \eq{Liepsie-2}, we can arrive  
\be\label{bJpsi}
\begin{split}
\bTh^{(3)}
&= -\ove{3!}\bep_{\m\r_1\r_2\r_3} \xi_\a \bar\Sigma^{\a\m} dx^{\r_1\r_2\r_3} 
	+ \ove{4!}  \bep_{\m\r_1\r_2\r_3} \bar\psi\g^{\a\b\m}\psi  \xi^\n \om_{\n\a\b}  dx^{\r_1\r_2\r_3} \\
&\quad~
	- \ove{3!}\frac{\eta_t-1}{4} \bep_{\m\r_1\r_2\r_3}   
		\xi^\a S_{\a\b\g} \,\bar\psi\g^{\b\g\m}\psi \,dx^{\r_1\r_2\r_3}\;.
\end{split}
\ee

Then, combining all the above with
\be
\xi\cdot\bL_R
= \xi\cdot \bep (R \,-\,2\L) 
= \ove{3!} \bep_{\m\r_1\r_2\r_3} \xi^\n \d^\m_\n (R \,-\,2\L) dx^{\r_1\r_2\r_3}, 
\ee
and using \eq{identities-4}, we can then rewrite the Noether current \eq{NoetherJ} into the following
\be   \label{bJetapen}
\begin{split}
\bJ_\xi 
&\approx 
	 \ove{3!\k^2} \bep_{\m\r_1\r_2\r_3} 
		\lrbrk{ (G^{\m\n} + \L g^{\m\n}) \xi_\n 
			 - \k^2 \xi_\n\bar\Sigma^{\n\m}
			 + \frac{\eta_t+1}{\eta_t}S^{\m\b\g}\nb_\g\xi_\b}  dx^{\r_1\r_2\r_3}  \\
&\quad 
	+ d\lrbrk{\ove{2\k^2}\ove{2!} \bep_{\m\b\r_2\r_3} \nb^\b\xi^\m dx^{\r_2\r_3}} 
	- \ove{3!}\frac{\eta_t+1}{4} \bep_{\m\r_1\r_2\r_3} 
		\xi^\a S_{\a\b\g}\, \bar\psi\g^{\b\g\m}\psi \, dx^{\r_1\r_2\r_3}
\end{split}
\ee

Let us `integrate by part' the last term in the big round bracket in the first line of \eq{bJetapen}, and using \eq{EOMS2nd} and \eq{identities-3} we can arrive
\be
\begin{split}
\ove{3!\k^2} \frac{\eta_t+1}{\eta_t} \bep_{\m\r_1\r_2\r_3} S^{\m\b\g}\nb_\g\xi_\b 
&=	d\lrbrk{ - \frac{\eta_t+1}{2\eta_t} 
			\ove{{\color{black}2}\k^2} \bep_{\m\b\r_2\r_3} S^{\m\b\g}\xi_\g 
			dx^{\r_2\r_3} }   \\
&\quad 
	- \frac{\eta_t+1}{2\eta_t}
		\ove{{\color{black}3}\k^2} S_{\b\m}{}^\s \bep_{\s\r_1\r_2\r_3} 
		S^{\m\b\g}\xi_\g dx^{\r_1\r_2\r_3}   \\
&\quad
	- \frac{\eta_t+1}{2\eta_t}
		\ove{{\color{black}3}} \bep_{\m\r_1\r_2\r_3} \Sigma^{[\m\n]}\xi_\n 
		dx^{\r_1\r_2\r_3}, 
\end{split}
\ee
where we have omitted the term proportional to $S_{\b\s}{}^\s$ which vanishes on-shell. Put it back to the Noether current expression \eqref{bJetapen}, 
we see that the quadratic torsion terms cancel each other, 
and the stress tensor part combined together well, i.e.,
\be
\begin{split}
-\k^2\xi_\n\bar\S^{\n\m}
-\k^2 (\eta_t+1) \bar\S^{[\m\n]}\xi_\n
&= -\k^2 \lrbrk{ \xi_\n\bar\S^{(\m\n)} + \eta_t \xi_\n\bar\S^{[\m\n]}} .
\end{split}
\ee  
Thus, we indeed obtain the graviton field equation in the big round bracket in the first line of \eqref{bJetapen} after `integrating by part'. Moreover, by imposing the on-shell conditions, the $\eta_t$ dependence is dropped and the Noether current is put into the desired form of \eq{bQEC}. 
 
   In summary, we have derived the explicit form of the symplectic potential \eq{bThEC} and the Noether charge \eq{bQEC} associated with a vector field $\xi$ for the Einstein-Cartan-fermion system. We can then use them to evaluate the corresponding quasi-local energy \eq{DeltaqE} or its variation \eq{qEsec2}.

    In the remaining of this paper we will apply the above results to some specific examples. One example is to evaluate the quasi-local energy for the perturbative solutions of AdS space due to the fermion field up to second order of Newton constant, and find that it agrees with the holographic relative entropy. The other example is to evaluate ADM mass as well as angular momentum for the asymptotically flat and AdS backgrounds.
 
\subsection{Comparison with holographic relative entropy}   \label{sec:eva}

 In this section we would like to explicitly evaluate the quasi-local energy \eq{qEsec2} for the perturbative solution, i.e., \eq{fmode} and \eq{backreactg}, of Einstein-Cartan-fermion system. We will find that our result coincides with the holographic relative entropy calculated in \cite{Lin:2016fua}.
 
  We will consider the quasi-local energy associated with the so-called entanglement wedge $\S$, which is a spatial region bounded by a boundary disk $B$ and the Ryu-Takayanagi minimal surface $\Bt$, whose area gives the holographic entanglement entropy associated with region $B$ for the dual CFT.  In \cite{Lashkari:2015hha,Lashkari:2016idm} it was argued that the quasi-local energy associated with a vector field satisfying the boundary conditions \eq{xionB}, \eq{binormal-c} with $\kappa_s=2\pi$, \eq{xibcsec2} and \eq{add-bc}\footnote{In \cite{Lashkari:2016idm} this condition is put in the asymptotical manner: $\cL_\xi g_{\m\n} \big|_{z\rightarrow 0}={\cal O}(z^{d-2})$ where $z$ is the radial coordinate of the AdS$_{d+1}$. This rapid falling behavior ensures that the modular Hamiltonian only receives the contribution from leading order perturbation.} is nothing but the holographic relative entropy of the dual CFT \footnote{The relative entropy for comparing reduced density matrices $\rho_A$ and $\sigma_A$ on region $A$ can be evaluated as follows: $S(\rho_A||\sigma_A)=\Delta\langle H_A\rangle -\Delta S_A$, where $H_A$ is the modular Hamiltonian and  $S_A$ is the entanglement. Here $\Delta$ means taking difference w.r.t. two different states. Holographically, $\langle H_A\rangle$ can be obtained via holographic stress tensor if region $A$ is a disk, and $S_A$ via Ryu-Takayanagi formula.}. Although the argument is very general, only some case of Einstein gravity is explicitly checked. Especially, in \cite{Lashkari:2015hha} the case is calculated in the Hollands-Wald gauge \cite{Hollands:2012sf} which requires the gauge transformation to fix the boundary conditions for the associated Killing vector field and the position of  $\Bt$.

  On the other hand, in \cite{Lin:2016fua} we evaluate the holographic relative entropy for the solution  \eq{fmode} and \eq{backreactg} of Einstein-Cartan-fermion system, and find that the resultant holographic relative entropy can be negative. If the equivalence between quasi-local energy and holographic relative entropy holds also for the Eisntein-Cartan-Fermion system, this implies that the positive energy condition can be violated. This is intriguing as there is no obvious pathology for the underlying theory and its solutions. Moreover, in this calculation we do not need to specify any vector field as in the defining the quasi-local energy. Therefore, it is not so trivial to check if the equivalence still holds for the gravity theory with torsion and fermion by direct evlaution of the quasi-local energy \eq{DeltaqE} for some vector field $\xi$ satisfying the aforementioned boundary conditions. 
  
\subsubsection{Effect of torsion and fermion}

  From the results of the symplectic potential \eq{bThEC} and the Noether charge \eq{bQEC}, it is obvious that there will be no direct torsion contribution to the physical charges. However, we emphasise that torsion can still contribute to the physical charges indirectly by sourcing the graviton. For example, the value of the quasi-local energy, which will be computed shortly, depends on $\eta_t$ explicitly. 
  
     Regarding the evaluation of the quasi-local energy based on the symplectic potential \eq{bThEC} and the Noether charge \eq{bQEC}, the first question is whether the additional terms related to fermion will contribute or not. We first consider the integral of \eq{qEsec2} over $\Bt$: in this case the vector field $\xi$ vanishes, i.e., \eq{xibcsec2} so that the $\xi\cdot \bTh$ and the second term in \eq{bQEC} vanish, too. Thus, only the usual term in Einstein gravity contributes to the integral of \eq{qEsec2} over $\Bt$.  
     
     On the other hand, the integral of \eq{qEsec2} over $B$ is more subtle: the second term in \eq{bQEC} will not contribute because its pullback vanishes on $B$ due to $\xi$ is time-like there. Thus, only the usual term of Einstein gravity in \eq{bQEC} contributes. For $\xi \cdot \bTh$ we notice that the second and third terms are sub-leading terms by $\k^2$ order in comparison with the first term, i.e., the usual term in Einstein gravity. This means that these terms could be suppressed by positive power of $z$ when approaching the boundary.  As we have the explicit solution \eq{fmode} and \eq{backreactg}, we can then perform the power counting of $z$ for those terms associated with fermion. After doing this, it turns out that both terms vanish on $B$ \footnote{By explicit check we can also see that the second term in $\xi\cdot \bTh$ vanishes by the tensor structure itself even before taking $z\rightarrow 0$.}, e.g., power counting of the third term of \eq{bThEC}:
\be
\sqrt{-g}(\d\bar\psi\g^\m\psi - \bar\psi\g^\m\d\psi)
= 
\frac{\ell^2}{z^4}
	\d^\m_z \frac{i \alpha ^2 \beta ^2 {r_L}^2 z^7 
				\left(3 {\eta_t}^2+2 {\mu_0} m^2 \ell ^2\right)}{3 \ell ^{12}} \sk
~\rar~ 0 \quad \text{as}\quad z\rar0. 
\ee

      From the above analysis, although the symplectic potential and Noether charge contain the terms associated with fermion, they however, will not contribute to the quasi-local energy  \eq{qEsec2}. Therefore, the effect of torsion and fermion comes into play only through the solution of field equations. We should say that this conclusion is quite general because the power counting is controlled by the fall-off behaviors of the on-shell solution, which however, is completely determined by the metric of AdS space. 

\subsubsection{Fix the vector field}

     The next step to evaluate the quasi-local energy is to choose appropriate vector field $\xi$ which satisfies the required boundary conditions. As mentioned, in \cite{Lashkari:2015hha} this was done by choosing Hollands-Wald gauge so that $\xi$ and positions of $B$ and $\Bt$ are fixed. This will save the efforts for finding the new $\xi$ in the deformed background but requires to solve the gauge conditions. The latter turns out to be a tedious procedure as demonstrated in \cite{Lashkari:2015hha}.  Instead, we will directly find the deformed vector field in the perturbative AdS space up to oder $\sk^2$, and then use it to evaluate the quasi-local energy. This should be equivalent to the evaluation in the Hollands-Wald gauge. 
     
     We start with the Killing vector of AdS space 
\be
\xi_{\text{AdS}} = \frac{\pi}{R_A} \lrbrk{ R_A^2-z^2 - r^2 - t^2 } \pa_t 
	- \frac{2\pi}{R_A} t ( z \pa_z 
					    	+ r \pa_r ) 
\ee
which satisfies the required conditions, and is the vector field used in the Hollands-Wald gauge. However, after the perturbation away from the pure AdS space there could be no Killing vector field. Despite that, for our purpose of evaluating quasi-local energy it is suffice to find a vector field that satisfies these boundary conditions on $B,\Bt$. In general, there are more than one solutions of $\xi$ that satisfy the above boundary conditions. However, the details of $\xi$ in the bulk of $\S$ is not relevant as the quasi-local energy \eq{qEsec2} is given by the integral over $B$ and $\Bt$. Therefore, all the $\xi$'s satisfying the same boundary conditions will yield the same quasi-local energy. 

    Since we solve the field equation perturbatively up to order $\sk^2$, we also only need to solve $\xi$ perturbatively up to the same order of $\sk$. Here, we will solve the solution by the following ansatz: 
\be \label{xi}
\xi = \frac{\pi}{R} \lrbrk{ r(z)^2 - r^2 - t^2 } \pa_t 
	- \frac{2\pi}{R}t \Big[ \lrbrk{f_0(z) + \sk f_1(z) + \sk^2f_2(z)}\pa_z 
					    + 
					    	g(r)\pa_r \Big], 
\ee
where $f_0 = z$, and 
\be
g(r) = \sqrt{R_A^2-z^2} + \sk g_1(z) + \sk^2g_2(z) + O(\sk^3)  \qquad\text{on $\Bt$}. 
\ee
Note also that $r(z)$ is the Ryu-Takayanagi minimal surface solved with respect to the perturbed metric \eq{backreactg}, and the details can be found in eq.(105)-(110) of \cite{Lin:2016fua}.

One finds that the real challenge of this ansatz is to solve the condition \eq{binormal-c}, 
and this will fix the unknown functions $f_1,f_2$ and $g_1,g_2$.  The explicit form of $f_1,f_2$ and $g_1,g_2$ is shown in Appendix \ref{app:xi}. Note that the second order functions $f_2$ and $g_2$ are quite complicated in comparison with $f_1$ and $g_1$. The former even contains the log pieces, which however is necessary to cancel the log pieces arising from the integral of the terms involving $f_1$ and $g_1$. In this sense, it is quite nontrivial to have the quasi-local energy evaluated here agree with the holographic relative entropy obtained in \cite{Lin:2016fua}, as shown below.

\subsubsection{Evaluation of the quasi-local energy}

  We are now ready to explicitly evaluate the quasi-local energy \eq{DeltaqE}  perturbatively up to the second order in $\sk$ based on the above discussions. Especially, all the terms involving the fermion in the symplectic potential and Noether charge will not contribute. Moreover, asymptotically, both the Noether charge as well as $-\xi\cdot\bK$ do not contribute beyond the linear order because of the asymptotic fall-off behaviours of fields.

    We first consider the integral over $\Bt$, for which only the first term in \eq{bQEC} contributes. Using the vector field \eqref{xi} with solutions \eqref{f1g1},\eqref{f2},\eqref{g2} found in the Appendix \ref{app:xi},  and the minimal surface $r(z)$ solved in \cite{Lin:2016fua}, we obtain
\be  \label{DeltaQBt}
\Delta \int_{\Bt} \bQ_\xi
= S_1+S_2, 
\ee
where 
\be  \label{SEEintBt}
\begin{split}
S_1
&= \frac{\pi ^2 \alpha  \beta  m{\mu_0} {R_A}^3}{2 \ell ^2}, \\
S_2
&= \frac{4 \pi ^3 \alpha ^2 \beta ^2 G_N R_A^6 \left(2 \eta_t^2-\mu_0^2 m^2 \ell ^2\right)}{35 \ell ^8}. 
\end{split}
\ee
Here $S_1$ is the first order result in $\sk$ and $S_2$ the second oder one. 
These are exactly eq.(113) and eq.(114) of \cite{Lin:2016fua}.  Although the functions $r(z)$, $f_2$ and $g_2$ are very complicated and contain log pieces, it is amazing that the log pieces all cancelled out to yield a simple final result.

   Now we consider the integral over the boundary disk $B$, on which  the vector $\xi$ reduces to the conformal Killing vector $\zeta$
\be
\xi \big|_{B} = \frac{\pi}{R_A}(R^2_A - r^2)\pa_t. 
\ee
 
   Unlike $\Bt$, the shape of $B$ is independent of the perturbation of the metric. Thus,  the integral of the Noether charge can be evaluated as follows and turns out to be 
\be  \label{QECintonB}
\begin{split}
\Delta \int_B \bQ_\xi=\int_B\bQ_\xi(\phi) - \int_B\bQ_\xi(\phi_0)= \frac{\pi ^2 \alpha  \beta  (3 {\mu_0}-4) m {R_A}^3}{12 \ell ^2}, 
\end{split}
\ee
while the integral of the symplectic potential term gives 
\be \label{ThetaECintonB}
\begin{split}
-\int_B\xi\cdot\bTh^\tR
&= \frac{\pi ^2 \alpha  \beta  (3 {\mu_0}+4) m {R_A}^3}{12 \ell ^2}. 
\end{split}
\ee
The use of symplectic potential suffices because $\bK$ does not contribute beyond linear order.

Sum up \eqref{QECintonB} and \eqref{ThetaECintonB}, we get 
\be\label{dHB}
\Delta \int_B  \lrbrk{ \bQ_\xi -  \xi\cdot\bK }
= \frac{\pi ^2 \alpha  \beta  {\mu_0} m {R_A}^3}{2 \ell ^2}, 
\ee
which agrees with the change of the expectation values of the modular Hamiltonian 
previously found in \cite{Lin:2016fua}.  Note that the above result is only first order in $\sk$ because of the fall-off behavior in $z$ of the perturbative solution for field equations. This kind of fall-off behavior also ensure the integrability condition \eq{intcond} and thus the existence of full $H_{\xi}$ as discussed earlier. 
 
    In summary, subtracting  \eq{DeltaQBt} from \eq{dHB}\footnote{The relative sign reflects the opposite normal directions of $B$ and $\Bt$.} we obtain the quasi-local energy for the perturbative solution \eq{fmode} and \eq{backreactg}:
 \be
 \D H_{\xi}=\frac{4 \pi ^3 \alpha ^2 \beta ^2 G_N R_A^6 \left(\mu_0^2 m^2 \ell ^2-2 \eta_t^2  \right)}{35 \ell ^8}.
 \ee   
This happens to be 
the same as the holographic relative entropy obtained in \cite{Lin:2016fua}. The positive energy condition can be violated if 
\be
\mu_0^2 m^2 \ell ^2 < 2 \eta_t^2\;.
\ee

\subsection{Fermion and torsion effect on ADM mass and angular momentum}   \label{sec:ADM}
Having the expression of quasi-local energy at hand, one is able to obtain the ADM mass or other global charges by simply replacing the spatial region $\S$ with the 2-sphere at infinity. We have seen from \eq{bThEC} and \eq{bQEC} that there are no direct contributions from torsion to the physical charges, as they are formally independent of $\eta_t$. However, we emphasise again that torsion effect can show up through sourcing the graviton field. With this observation, it is then natural to ask if the presence of fermion deforms the global charges. As far as we know, the inclusion of fermions is rarely discussed in the literature. We will answer this question for the ADM mass and angular momentum in both asymptotically flat space-time as well as AdS background. 
  
\subsubsection{Review of ADM quantities in Einstein gravity}
We first review the derivation of ADM mass in asymptotically flat space-time for pure Einstein gravity as done in \cite{Iyer:1994ys}, to demonstrate how the quasi-local energy is linked to the ADM mass. The variation of the ADM mass is given by 
\be  \label{ADMsec5}
\d H_\xi = \int_{S^2_{\infty}} \lrbrk{\d\bQ_\xi - \xi\cdot\bTh}, 
\ee
where $S^2_{\infty}$ is the 2-sphere at infinity and $\xi$ is the asymptotic Killing vector $\xi :=  \hat t = \pa_t$. To find the ADM mass, we need to find out $\bK$ defined in \eq{defbK} and set the absolute value of $H_{\hat t}$ for Minkowski space as zero. 
An asymptotically flat space-time has the following boundary condition of metric 
\be  \label{asympFlat}
g_{\m\n} = \eta_{\m\n} + O(1/r), 
\ee
where $\eta_{\m\n}$ is the metric of Minkowski space, and $r$ is the radial component in the polar coordinates 
\be  \label{polarAsympflat}
ds^2 = -dt^2 + dr^2 + r^2 d\th^2 + r^2\sin^2\th d\varphi^2 + O(1/r). 
\ee
To find $\bK$, we first evaluate $- \hat{t} \cdot\bTh$ and try to pull out $\d$ in accordance with \eq{asympFlat}. The result is
\be
\int_{S^2_{\infty}} \hat{t} \cdot \bTh = 
	-\ove{2\k^2} \d \int_{S^2_{\infty}} dS 
		\lrbrk{ (\pa_r g_{00} - \pa_0 g_{r0})
			   + r^k h^{ij} (\pa_i h_{kj} - \pa_k h_{ij}) }, 
\ee
where $i,j,k=1,2,3$, $r^k = \d^k_r$, $h_{ij}$ is the spatial metric, $dS$ is the area element of 2-sphere of radius $r$, and we have repeatedly used the asymptotic boundary condition \eqref{asympFlat}. $\tilde{\bep}_{\m\n\r\s}$ is the Levi-Civita symbol and $\tilde{\bep}_{0r\th\varphi} = 1$. 
Therefore, $\bK$ is given by 
\be  \label{bKADM}
(\hat{t} \cdot\bK)_{\a\b}
= -\ove{2\k^2} \bep_{\a\b} 
	\lrbrk{ (\pa_r g_{00} - \pa_0g_{r0}) + r^kh^{ij}(\pa_i h_{kj} - \pa_k h_{ij}) }. 
\ee
Similarly, the Noether charge part can be calculated straightforwardly, 
\be  \label{bQADM}
\int_{S^2_{\infty}} \bQ_{\hat t} =
	 -\ove{2\k^2} \int_{S^2_{\infty}} dS
	 	 \lrbrk{\pa_r g_{00} - \pa_0g_{r0}}, 
\ee
where the asymptotic boundary condition \eq{asympFlat} is used in the last equality. 
Notice that the higher sub-leading terms in the asymptotic expansion of metric does not contribute. 

Sum up \eq{bQADM} and the integration of \eq{bKADM}, we get precisely the well-known ADM mass formula 
\be
\begin{split}
H_{\hat t}
&= \int_{S^2_{\infty}} \lrbrk{\bQ_{\hat t} - \hat{t} \cdot\bK} 
= \ove{2\k^2} \int_{\infty} dS \, r^k h^{ij}(\pa_ih_{kj} - \pa_k h_{ij})  
:= M_{\ADM}. 
\end{split} 
\ee

If the system admits asymptotic rotational Killing vector, e.g. $\hat{\varphi} := \ove{r\sin\th}\frac{\pa}{\pa\varphi}$, there is an associated angular momentum defined by\footnote{The minus sign comes from the fact that the Killling vector $\hat{\varphi}$ is space-like.}
\be  \label{angmom}
J_{\hat \varphi} := - \int_{S^2_{\infty}} \bQ_{\hat \varphi}.  
\ee
Notice that the $-{\hat \varphi} \cdot\bK$ term drops because it is pullback to zero on $S^2_{\infty}$. 
For the Einstein gravity, this form of angular momentum is exactly in the form of the Komar formula \cite{Komar:1958wp}\footnote{It is worth noting that the Komar anomalous factor of 2 between Komar mass and angular momentum formulas is naturally resolved in Wald's formalism \cite{Iyer:1994ys}. }, 
\be
J_{\hat{\varphi}}^{(\text{Komar})}
= \ove{2!}\ove{2\k^2} \int_{S^2_{\infty}} \bep_{\a\b\r_1\r_2} \tilde{\nb}^{\a}\xi^\b dx^{\r_1\r_2}. 
\ee

\bigskip

\subsubsection{ADM quantities in Einstein-Cartan-fermion system}
Now, we are in a position to go beyond Einstein gravity and check if the fermion contributes to the ADM mass or angular momentum. 
We first consider again the asymptotically flat space-time. 
As we will not find out $\bK$ explicitly in the torsion gravity case, we need to confirm its existence by the integrability condition \eq{intcond}. To this end, we compute the symplectic current, 
\be  \label{sympcurrent}
\begin{split}
\bo
&= \ove{2\k^2}\bep_{\a} P^{\a\b\m\n\r\s}
	\lrbrk{\d_2 g_{\b\m} \tilde\nb_{\n} \d_1 g_{\r\s} 
		   - \d_1 g_{\b\m} \tilde\nb_{\n} \d_2 g_{\r\s}},   \\
&\quad 
	+ \bep_\m 
		\bigg[ -\ove{4}(\bar\psi\g^{\a\b\m}\psi) e^\g_b e_{\b a} \d_1e^b_\g \d_2e^a_\a  
				\\
&\qquad\quad~~
			  - \ove{4} \d_1(\bar\psi\g^{\a\b\m}\psi) \d_2e^a_\a \,e_{\b a}   \\
&\qquad\quad~~
			  - \ove{4} (\bar\psi\g^{\a\b\m}\psi) \d_2 e^a_\a \d_1 e_{\b a}  \\
&\qquad\quad~~
			  + \ove{2} \lrbrk{\d_2\bar\psi \g^\m\psi - \bar\psi\g^\m\d_2\psi} 
			  	e^\a_a \d_1e^a_\a    \\
&\qquad\quad~~
			  + \ove{2} \lrbrk{ \d_2\bar\psi\g^\m \d_1\psi - \d_1\bar\psi\g^\m\d_2\psi } \\
&\qquad\quad~~
			  - \ove{2} \lrbrk{\d_2\bar\psi\g^\a\psi - \bar\psi\g^\a\d_2\psi}
			  	e_b^\m \d_1 e^b_\a
		\bigg] - (\d_1 \lrar \d_2), 
\end{split}
\ee
where $-(\d_1\lrar\d_2)$ is only applied to the second term above, and 
\be
P^{\a\b\m\n\r\s}
= g^{\a \r}g^{\s \b}g^{\m\n} 
	- \ove{2}g^{\a \n}g^{\b \r}g^{\s \m} 
	- \ove{2}g^{\a\b}g^{\m\n}g^{\r\s}
	- \ove{2}g^{\b \m}g^{\a \r}g^{\s \n} 
	+ \ove{2}g^{\b \m}g^{\a \n}g^{\r\s}. 
\ee
Remind that $\tilde{\nb}_\m$ is the Riemannian covariant derivative without including torsion.

To satisfy \eq{intcond}, with \eq{asympFlat}, we need the following asymptotic boundary condition of fermion 
\be  \label{psidpsifalloff}
\d\psi \sim \ove{r^n}, \quad n >1, \qquad\qquad \psi \sim \ove{r^m}, \quad m>0. 
\ee
In fact, we can do better. 
As $\d\f$ satisfies the linearized field equations in Wald's formalism,  their asymptotic boundary conditions are constrained by on-shell relations. Hence, we may start from the marginal value of the boundary condition for metric, i.e. the value that gives finite ADM mass, $g = \eta + O(1/r)$, and look for the boundary condition of fermions by the consistency of field equation. 

The Minkowski background version of the perturbative Einstein field equation for Einstein-Cartan-fermion system is given by \cite{Lin:2016fua}
\be\label{linear-1}
\Gt^\1_{\m\n}  =   \st^\0_{\m\n}. 
\ee
whose linearized version is, schematically \footnote{In the RHS of \eq{linear-1} and \eq{perbEinsteinEOM} appearing in \cite{Lin:2016fua} there is an overall IR factor $r_L^2$. For simplicity, we just set it to one here.}, 
\be  \label{perbEinsteinEOM}
\square \d g_{\m\n}  
=  \tilde\nb_{(\m}^\0\bar\psi\g^\0_{\n)}\psi^\0 
					- \bar\psi\g^\0_{(\n}\tilde\nb_{\m)}^\0\psi^\0, 
\ee
where $\square$ denotes a second-order derivative operator that includes the Laplacian. 
As $\d g \sim 1/r$, by simple power counting of \eq{perbEinsteinEOM} we conclude that 
\be  \label{frommarginalgMin}
\psi^\0 \sim 1/r\;.
\ee
Notice that \eq{frommarginalgMin} is compatible with \eq{psidpsifalloff} as $\d\psi$ should be sub-leading in $1/r$ expansion. Thus, $\d\psi$ should obey \eq{psidpsifalloff} so that the integrability condition is indeed satisfied. 

   Similarly, further power counting can tell us if the non-Einstein parts of the Noether charge \eq{bQEC} and the symplectic potential \eq{bThEC} will contribute to the ADM mass and angular momentum or not. 
A straightforward power counting then shows that all the fermion terms in $-\xi\cdot\bTh$ do not contribute to the ADM mass. For the fermion part in the Noether charge, notice that it is pullback to zero on $S^2_{\infty}$ due to the tensorial structure and $\xi = \hat{t}$. Therefore, we find that the fermion does not contribute to the ADM mass for our Einstein-Cartan-fermion system in asymptotically flat space. 

However, it is a different story for angular momentum \eqref{angmom}, because the fermion term in the Noether charge no longer has vanishing pullback to $S^2_{\infty}$. This is so because the Killing vector now is  $\hat \varphi$ not $\hat t$. Moreover, the power counting shows that the fermion part has a finite contribution. Therefore, we expect the presence of the spin 1/2 fermion deforms the angular momentum and this may be an observable physical effect such as in the compact binary inspirals \cite{Steinhoff:2008zr,Steinhoff:2008ji,Steinhoff:2014kwa}. 
Let us perform the power counting explicitly for this finite contribution. 
Use the metric \eq{polarAsympflat}, we construct vielbeins connecting the polar coordinates with the orthonormal frame: 
\be
{\bf e}^{(t)} = dt, \qquad {\bf e}^{(r)} = dr, \qquad 
{\bf e}^{(\th)} = rd\th, \qquad {\bf e}^{(\varphi)} = r\sin\th\, d\varphi, 
\ee
where the boldface letter ${\bf e}$ again represents differential forms, 
and we have used the parenthesised indices to label the orthonormal frame indices. 
Notice that, in particular, 
\be  \label{gavarphiMin}
{\bf e}^{(\varphi)}_\varphi = r\sin\th, \qquad\Rightarrow\qquad 
\g^\varphi = g^{\n\varphi} e^a{}_{\n} \g_a
= g^{\varphi\varphi} e^{(\varphi)}_\varphi \g_{(\varphi)}
= \ove{r\sin\th}\g_{(\varphi)}. 
\ee
Now, let us check the contribution of the fermion term in \eqref{angmom},  
\be
\begin{split}
\bep_{\a\b\r_1\r_2} \bar\psi\g^{\a\b\g}\psi\,  g_{\g\d}\xi^\d dx^{\r_1\r_2}  
=
	4 \bar\psi\g^{0r\varphi}\psi\, (r^2\sin^2\th) \ove{r\sin\th} 
		(r^2\sin\th\,d\th d\phi) 
\end{split}
\ee
which is finite because $\bar\psi \sim \psi \sim 1/r$ and $\g^\varphi \sim 1/r$ according to \eq{gavarphiMin}. Therefore, when the fermion is present, we find that the angular momentum is the following extension of the Komar formula, 
\be \label{angmomKomarlike}
\begin{split}
J_{\hat{\varphi}}
&= J_{\hat\varphi}^{(\text{Komar})} 
- \ove{2!}\ove{2} 
	\int_{S^2_{\infty}} \bep_{\a\b\r_1\r_2}  
		\ove{4}\bar\psi\g^{\a\b\g}\psi \, \xi_\g dx^{\r_1\r_2},   \\
&= J_{\hat\varphi}^{(\text{Komar})} 
	-\ove{4} \int_{S^2_{\infty}} \bar\psi\g_{\r_2}\g_5\psi \,\xi_{\r_1} dx^{\r_1\r_2}, 
\end{split}
\ee
where $J_{\hat\varphi}^{(\text{Komar})}$ is the Komar angular momentum and we follow the gamma matrix convention in \cite{Lin:2016fua}. 
This means the axial current generates angular momentum for Einstein-Cartan-fermion system in asymptotically flat space-time.

For the case of AdS background, we have explicit solutions \eq{fmode} and \eq{backreactg}. It is then straightforward to do power counting, and conclude that the existence of fermions does not deform ADM mass or angular momentum formulas. 

\section{Implications to gravitational wave physics}  \label{sec 4}

  LIGO's discoveries of the gravitational waves from binary black hole inspirals and mergers have opened an new era of gravitational wave astronomy. Up to this moment, there are already three events \cite{Abbott:2017vtc, Abbott:2016nmj, TheLIGOScientific:2016src} confirmed as binary black hole mergers, and we may expect dozen or hundred to come in the near future. As  the more and more data of gravitational wave are collected, the more precision test of Einstein gravity is expected. Therefore, this is the right timing to push the modified gravity theories into the regime of precision test. As the Einstein gravity has passed many precision tests in the scale of solar system, e.g. \cite{Bertotti:2003rm, Shapiro:1964uw}, we will not expect there are detectable deviations from Einstein gravity in the weak field regime. On the other hand, black hole mergers of all the LIGO discovery events up to this moment are in the strong field regime, for which we may like to test the deviation from Einstein gravity in a more precise way. Some constraints on modified gravities derived from the LIGO's discoveries are done, for example, in \cite{Yunes:2016jcc}, which however, does not include torsion gravity. 
  
  There are many proposals of modified gravity theories, the most common one is the scalar-tensor gravities such as the Brans-Dickie theory, or more general Hordenski theory. They can be treated formally in the usual way of second order metric formulation of Einstein gravity. On the other hand, by adding the torsion, as shown in this work, one needs to adopt the first order formulation involving veilbeins and spin connections. 
  
   There are many physical reasons to introduce torsion into play, and now also in the new gravitational wave physics. The key point is that the torsion is naturally sourced by some high energy coherent states of fermionic matters, which are the main constituents of all the astronomical objects including black holes and neutron stars.  We can imagine that for massive fermion stars, say hundred of solar masses, the torsion coupling could affect the pattern of gravitational radiation originated from the inspirals and mergers of such stars. Especially, there are possibilities that some fermions are dark matter candidates, which form the dark fermion stars which could only couple to gravity. In this case the LIGO observations can serve as the window into the dark matters and the torsion gravity. Moreover, in order to find out the waveform of these dark stars' merger, we need to implement the numerical relativity calculations. The numerical relativity is formulated as the 3+1 Cauchy problem, e.g. the famous BSSNOK formulation \cite{Shibata:1995we, Nakamura:1987zz, Baumgarte:1998te}, which however, is formulated in terms of metric and extrinsic curvature, not of the veilbein and spin connection. To develop the similar 3+1 Cauchy problem for torsion gravity coupled to dark fermion matters, we instead need to adopt the first order formulation in terms of veilbein and spin connection, and the formulation developed in this work should be quite useful for such purpose. We should emphasize that dynamics of the dark stars should be clear, simpler and more massive than the neutron stars without the complication of the nuclear interactions and the Tolman-Oppenheimer-Volkoff limit on the mass of neutron stars. This shall be good for testing the modified gravity such as torsion gravity by the gravitational radiation in the strong field regime.
   
   On the other hand, even without introducing the fermion matters, the torsion could also be induced at the low energy by either the high energy fermion matters or the nonlinear self-interaction of gravitons, e.g., the Routhian for the low energy effective dynamics of the spin bodies of mass $m$ contains the terms
\be
 {1\over 2} \omega_{\m}^{ab} S_{ab} u^{\mu} + {1\over 2m} R_{\a\b\g\d}S^{\g\d}u^{\a}S^{\b\s}u_{\s}
\ee
where $S^{ab}$ is the spin tensor in the local flat frame characterized by the tetrad $e^a_{\m}$, $S^{ab}:=S^{\m\n}e^a_{\m}e^b_{\n}$, $u^{\m}$ is the 4-velocity of the spinning body. In the above,  the Riemann tensor $R_{\a\b\g\d}$ and the spin connection $\omega_{\m}^{ab}$ are the ones for torsion gravity defined in \eq{cpntRmnab} and \eq{spinconnection-torsion}, respectively. By using this Routhian with the help of the tricks developed in this work, we believe that one can derive the gravitational waveform of the inspirals of the binary black holes by following the effective approach to the post-Newtonian (PN) gravity developed in \cite{Goldberger:2004jt, Porto:2005ac, Levi:2014sba}.   With the incorporation of torsion in this way, its effect for the gravitational waveform should be comparable with the pure Riemannian gravity effect at higher PN order, e.g., 4PN.    Furthermore, this torsion effect could be significant for the self-force of a spin body around a supermassive Kerr black hole, which can also be studied by generalizing the standard self-force problem, e.g., \cite{Barack:2002ku}. The result should be relevant to the quasi-normal modes in the merger phase of binary black hole due to the torsion effect. We remark that the 1PN analysis in the context of Poincare gauge theory \cite{Hehl:1979xk} was already done more than three decades ago \cite{Schweizer:1980vn} and it was found no difference from the Einstein gravity. This is however expected as the torsion effect should manifest as the finite-size effect at the higher PN order. In summary, it is optimistic to see the torsion effect when the precision of detectors is improved in the future.

\section{Conclusion}\label{conclusion}

The positive energy condition is important on the issues of stability of solutions for the theory of gravity. However, it was mostly done for the Einstein gravity. In this paper we adopt Wald's formalism of covariant Noether charge to construct the explicit expression of quasi-local energy for the Einstein-Cartan-fermion system. We evaluate the quasi-local energy of entanglement wedge for some particular solution, and find that it is not always positive definite for arbitrary fermion mass and torsion-axial current coupling. This result implies the violation of the positive energy theorem for Einstein-Cartan-fermion system, and provide a nontrivial example of swampland beyond the symmetry principle and possibly the weak gravity conjecture. Though we cannot see a whole scope of the swampland by a simple example, it can be a step stone for further exploration of the issue. Moreover, the on-shell value of the quasi-local energy agrees with the holographic relative entropy evaluated in \cite{Lin:2016fua}. This generalize their equivalence from Einstein gravity to the torsion gravity. 

  Besides, we find that the quasi-local energy is \textit{formally} independent of the torsion-axial current coupling, although the dependence will come in through the solutions of the field equations when evaluating it on-shell. Despite that, this ``pseudo-topological" nature is unexpected and thus intriguing, and may have some deep implication for AdS/MERA duality along the similar line of bulk/edge correspondence for topological insulator. 

   Torsion is a very natural addition to Einstein gravity in the presence of fermion sources, and thus torsion gravity should be called for the tests of the observation data. In view of this, we have also discussed about the implication of our results to the tests of torsion gravity in the context of new gravitational wave astronomy. We elaborate that the techniques developed here should be useful in generating the gravitational waveform templates of torsion gravity from either effective-one-body formalism or from numerical relativity simulation, because there is not much discussions in the literatures on the first order formulation in terms of vielbein and spin connection in the aforementioned approaches. We are currently working along this direction, and hope to report the results in the near future.

\subsection*{Acknowledgements}
This work is partially supported by NCTS. FLL is supported by Taiwan Ministry of Science and Technology through Grant No.~103-2112-M-003 -001 -MY3, No.~103-2811-M-003 -024, No.~106-2112-M-003 -004 -MY3, No.~106-2918-I-003 -008, and No.~ 06-2811-M-003 -026. BN is supported in part by NSFC under Grant No.~11505119.  We thank Carlos Cardona, Chong-Sun Chu, Dimitrios Giataganas, Kyung Kiu Kim, Rong-Xin Miao, Pichet Vanichchapongjaroen,  Dong-han Yeom for their discussions.  This project is initiated at the IF-YITP GR+HEP+Cosmo International Symposium VI,  3rd-5th August, 2016. FLL thanks members of IF for the hospitality during his visit.  Sh-L. K. is grateful to NCTS and NTNU for their kind hospitality during his visits when working in progress.

\appendix

\section{Solving the vector field $\xi$ up to $O(\kappa^4)$}   \label{app:xi}

From the minimal surface equation 
\be
r = r(z), 
\ee
which is also obtained in \cite{Lin:2016fua}, 
one obtains the two unit normal vectors as 
\be
n_\1^\m = -\frac{g^{0\m}}{\sqrt{-g^{00}}}, \qquad  
n_\2^\m = \ove{\sqrt{g^{rr}-2g^{zr}r'(z) + g^{zz}r'(z)^2}}
			\lrbrk{g^{\m r} - g^{\m z}r'(z)}. 
\ee
The binormal vector is then 
\be
\begin{split}
n^{\m\n}
&= n_\1^\m n_\2^\n - n^\m_\2 n^\n_\1  \\
&= \ove{\sqrt{-g^{00}(g^{rr} - 2g^{zr}r' + g^{zz}r'^2)}}
	\lrsbrk{-g^{0\m}g^{r\n} + g^{0\m}g^{z\n}r'(z) - (\m\lrar\n)}, 
\end{split}
\ee
which has only nonzero (independent) components $(tr)$ and $(tz)$. 
On the other hand, 
\be
\tilde{\nb}^{[\m}\xi^{\n]}
= g^{\m[\r}g^{\s]\n} 
	\pa_\r\lrbrk{\xi^\t g_{\t\s}}. 
\ee
The boundary condition \eq{binormal-c}  then gives us two equations for the unknown functions at each order. One then obtains the following solutions: 

\be  \label{f1g1}
g_1 = \frac{\alpha  \beta  \mu_0 m r_L^2 z^3 \left(3 z^2-5 R_A^2\right)}{8 \ell ^4 \sqrt{R_A^2-z^2}}, \quad 
f_1 = -\frac{\alpha  \beta  \mu_0 m r_L^2 z^2 \left(7 z^2-3 R_A^2\right)}{8 \ell ^4}, 
\ee

\be  \label{g2}
\begin{split}
g_2(z)
&= \ove{{40320 R_A^2 \ell ^{10} \left(R_A^2-z^2\right)^{3/2}}}
 \alpha ^2 \beta ^2 r_L^4   \\
&\quad
 \bigg[7776 \mu_0^2 m^2 R_A^8 \ell ^2 
 		\left(R_A^4-3 R_A^2 z^2+2 z^4\right) \log (R_A)    \\
&\quad
	-7776 \mu_0^2 m^2 R_A^8 \ell ^2 
		\left(R_A^4-3 R_A^2 z^2+2 z^4\right) \log (R_A+z)  \\
&\quad	+z \Big(7776 \mu_0^2 m^2 R_A^{11} \ell ^2
			-3888 \mu_0^2 m^2 R_A^{10} z \ell ^2
			-7776 \mu_0^2 m^2 R_A^9 z^2 \ell ^2
			+4050 \mu_0^2 m^2 R_A^8 z^3 \ell ^2 	\\
&\qquad~~ 		+1485 \mu_0^2 m^2 R_A^6 z^5 \ell ^2
			+6720 \mu_0 m^2 R_A^6 z^5 \ell ^2
			-4480 m^2 R_A^6 z^5 \ell ^2
			+33993 \mu_0^2 m^2 R_A^4 z^7 \ell ^2  \\
&\qquad~~
			-13440 \mu_0 m^2 R_A^4 z^7 \ell ^2
			+8960 m^2 R_A^4 z^7 \ell ^2
			-77220 \mu_0^2 m^2 R_A^2 z^9 \ell ^2
			+6720 \mu_0 m^2 R_A^2 z^9 \ell ^2  \\
&\qquad~~		-4480 m^2 R_A^2 z^9 \ell ^2
			+40320 \mu_0^2 m^2 z^{11} \ell ^2
			+10080 \eta_t^2 R_A^4 z^7
			-20160 \eta_t^2 R_A^2 z^9
			+10080 \eta_t^2 z^{11}
	\Big)\bigg], 
\end{split}
\ee
\be  \label{f2}
\begin{split}
f_2(z)
&= \ove{{20160 R_A^2 \ell ^{10} (R_A+z)}}
\alpha ^2 \beta ^2 r_L^4 z   \\
&\quad
\bigg[ -7776 \mu_0^2 m^2 R_A^9 \ell ^2 \log (R_A+z)
	+7776 \mu_0^2 m^2 R_A^9 \ell ^2 \log (R_A)
	+3888 \mu_0^2 m^2 R_A^8 z \ell ^2   \\
&\quad~
	+7776 \mu_0^2 m^2 R_A^8 z \ell ^2 \log (R_A)
	-7776 \mu_0^2 m^2 R_A^8 z \ell ^2 \log (R_A+z)
	-4671 \mu_0^2 m^2 R_A^5 z^4 \ell ^2   \\
&\quad~ 
	-4671 \mu_0^2 m^2 R_A^4 z^5 \ell ^2
	-7110 \mu_0^2 m^2 R_A^3 z^6 \ell ^2
	+3360 \mu_0 m^2 R_A^3 z^6 \ell ^2
	-2240 m^2 R_A^3 z^6 \ell ^2  \\
&\quad~	
	-7110 \mu_0^2 m^2 R_A^2 z^7 \ell ^2
	+3360 \mu_0 m^2 R_A^2 z^7 \ell ^2
	-2240 m^2 R_A^2 z^7 \ell ^2
	+20160 \mu_0^2 m^2 R_A z^8 \ell ^2  \\
&\quad~ 
	+20160 \mu_0^2 m^2 z^9 \ell ^2
	-5040 \eta_t^2 R_A^3 z^6-5040 \eta_t^2 R_A^2 z^7
	+5040 \eta_t^2 R_A z^8+5040 \eta_t^2 z^9
\bigg]. 
\end{split}
\ee

\providecommand{\href}[2]{#2}\begingroup\raggedright\endgroup

\end{document}